\documentclass[12pt]{article}

\usepackage{times}
\usepackage{graphicx}
\usepackage{amssymb}
\usepackage{amsthm}
\usepackage{amsmath}
\usepackage{dsfont}
\usepackage{bm}
\usepackage{mathrsfs}
\usepackage{bbold}

\def\Xint#1{\mathchoice
{\XXint\displaystyle\textstyle{#1}}%
{\XXint\textstyle\scriptstyle{#1}}%
{\XXint\scriptstyle\scriptscriptstyle{#1}}%
{\XXint\scriptscriptstyle\scriptscriptstyle{#1}}%
\!\int}
\def\XXint#1#2#3{{\setbox0=\hbox{$#1{#2#3}{\int}$ }
\vcenter{\hbox{$#2#3$ }}\kern-.6\wd0}}

\def\dashint{\Xint-}

\begin{document}

\title{Lifshitz theory of the cosmological constant}
\author{Ulf Leonhardt\\
Department of Physics of Complex Systems,\\
Weizmann Institute of Science,\\ Rehovot 7610001, Israel
}
\date{\today}
\maketitle

\begin{abstract}
Astrophysics has given empirical evidence for the cosmological constant that accelerates the expansion of the universe. Atomic, Molecular, and Optical Physics has proven experimentally that the quantum vacuum exerts forces --- the van der Waals and Casimir forces --- on neutral matter. It has long been conjectured [Ya.~B. Zel'dovich, Usp. Fiz. Nauk {\bf 95}, 209 (1968)] that the two empirical facts, the cosmological constant and the Casimir force, have a common theoretical explanation, but all attempts of deriving both from a unified theory in quantitative detail have not been successful so far. In AMO Physics, Lifshitz theory has been the standard theoretical tool for describing the measured forces of the quantum vacuum. This paper develops a version of Lifshitz theory that also accounts for the electromagnetic contribution to the cosmological constant. Assuming that the other fields of the Standard Model behave similarly, gives a possible quantum--optical explanation for what has been called dark energy.\\

\end{abstract}

\newpage

\section{Argumentation}

\subsection{Introduction}

Einstein \cite{Einstein} introduced the cosmological constant $\Lambda$ for having the possibility of a static, eternal universe as solution of his field equations of gravity \cite{LL2}. The cosmological term he wrote there, acts as a repulsive force that may counter--balance the gravitational attraction of ordinary matter in equilibrium \cite{Peacock}. Hubble's astronomical observations \cite{Hubble} however, of galaxies receding from each other on average, revealed a different picture \cite{Harrison}: the universe is not static, cosmic distances are expanding with a universal, time--dependent factor. The first derivative of the expansion factor differs from zero and is positive. More recent measurements with supernova explosions  \cite{Supernovae1,Supernovae2} and of the Cosmic Microwave Background (CMB) \cite{CMBPlanck,CMB} have refined Hubble's results with sufficient precision to determine the second derivative of the expansion factor, that turned out to be positive, too: the expansion of the universe is accelerating. This is only possible if, on cosmological scales, the net force of gravity is repulsive, which gives strong, empirical evidence for the cosmological constant. The analysis of CMB fluctuations \cite{CMBPlanck} has established the currently best quantitative value of $\Lambda = (1.106 \pm 0.023)\times 10^{-18}\,\mathrm{m}^{-2}$ (valid for the time the CMB was formed). However, predictions of standard quantum field theory exceed the empirical value of $\Lambda$ by about 120 orders of magnitude \cite{Weinberg}. The problem arises from the nature of the quantum vacuum \cite{Milonni}.

The universe, with an average density \cite{CurveMeas} of approximately\footnote{To a good approximation, the universe is spatially flat \cite{CurveMeas} such that Friedmann's Eq.~(\ref{f1}) holds. Relating the mass density $\varrho$ to the energy density as $\epsilon/c^2$ gives the quoted approximate value of $\varrho$ for the Hubble constant $H=2.2\times 10^{-18}\mathrm{Hz}$ from CMB measurements \cite{CMBPlanck}.} $10^{-29}\, \mathrm{g}/\mathrm{cm}^3$, is mostly made of empty space, but this cosmic vacuum is thought to be filled with quantum fields in their ground state --- the quantum vacuum \cite{Milonni} that may cause the measured cosmological force \cite{Zeldovich}. Empirical evidence \cite{Rodriguez,KMM,Lambrecht,Levitation,CasimirEquilibrium,Decca} for forces of the quantum vacuum \cite{Forces} comes from Atomic, Molecular, and Optical (AMO) Physics. Here they appear \cite{Rodriguez} as the van der Waals \cite{Shahmoon} and Casimir forces \cite{Scheel}, and here they agree with theory up to an accuracy on the percent level that is only limited by the experimental precision of the material parameters involved  \cite{Rodriguez}, in contrast to cosmology. Zel'dovich \cite{Zeldovich} suggested that the quantum vacuum appears on cosmological scales as the cosmological constant. His theory \cite{Zeldovich} predicts the correct structure of the cosmological term, but a vastly incorrect quantitative value, and so did other theories \cite{Weinberg,DarkEnergy} or they could not account for the empirically observed forces of the quantum vacuum \cite{Caldwell,Other}. Perhaps for want of a more illuminating explanation, the cosmological constant has been called dark energy \cite{DarkEnergy}.

\begin{figure}[t]
\begin{center}
\includegraphics[width=16pc]{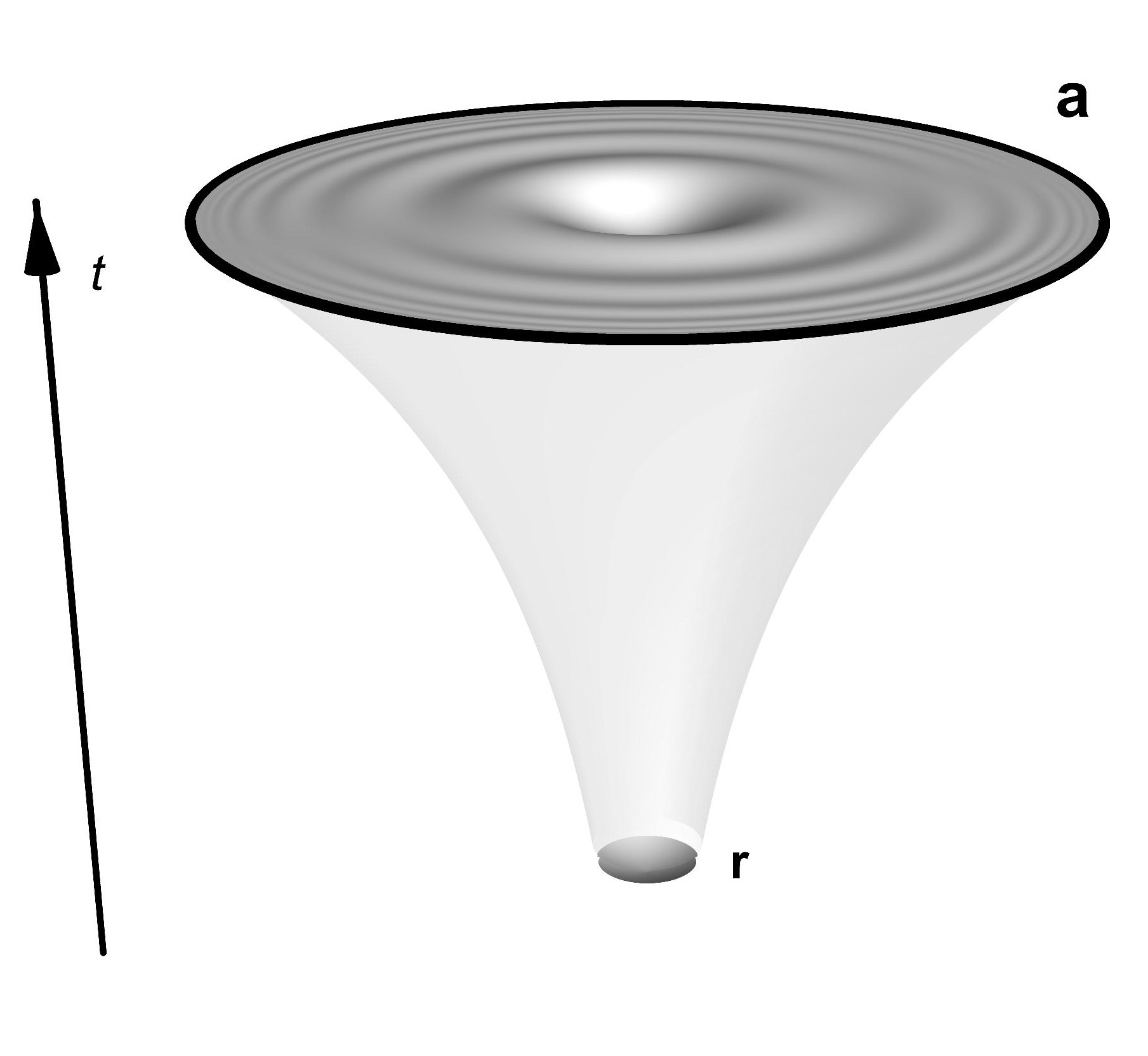}
\hspace{2pc}
\includegraphics[width=16pc]{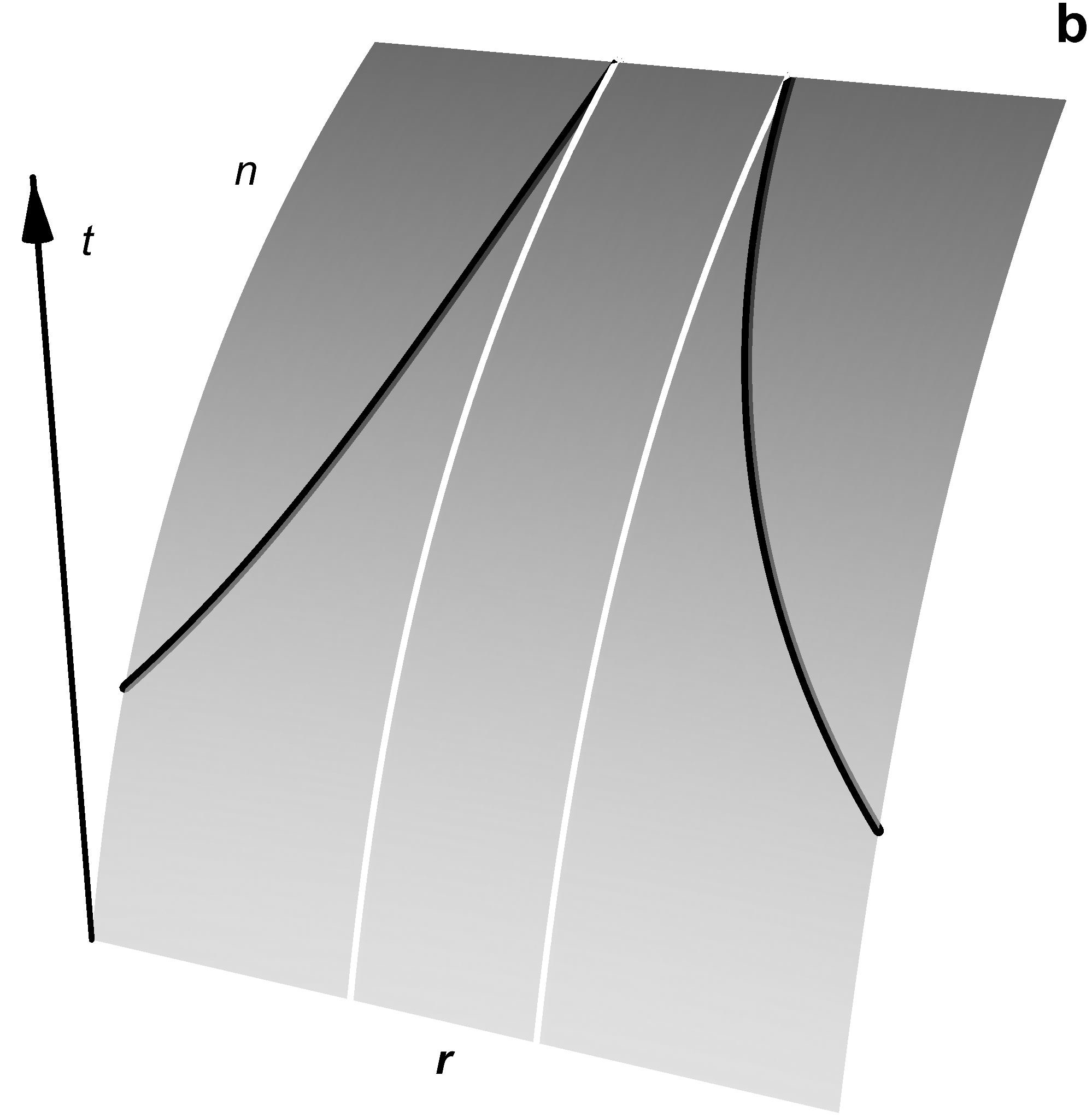}
\caption{
\small{
Cosmic expansion. {\bf a:} Visualization of the expansion of the universe in a space--time diagram; three--dimensional space is illustrated as a plane in coordinates $\bf{r}$ affixed to the origin, Eq.~(\ref{fix}), time $t$ appears as a third dimension. At the black circle the shown patch of space reaches a cosmological horizon where the expansion velocity equals the speed of light. No communication from beyond that sphere is possible. The waves show the last signal from the outside reaching an observer in the center (Appendix A). {\bf b:} Representation of the expanding universe as a medium with time--dependent refractive index $n(t)$, illustrated for one--dimensional space in co--moving coordinates. The shades of gray and the bending of the space--time sheet visualize the variation of $n$. The white lines mark, in co--moving coordinates, the boundary of the same patch shown in fixed coordinates in {\bf a}, reaching the horizon (black lines). 
}
\label{expansion}}
\end{center}
\end{figure}

Given the success \cite{Rodriguez} of the theory of the quantum vacuum in AMO Physics, it seems natural to take a similar approach for calculating the cosmological constant, which is what this paper strives to achieve. The starting point is the observation \cite{Gordon,Quan1,Quan2,Stor,Plebanski,Schleich,LeoPhil} that a space--time geometry appears as an effective medium to the electromagnetic field\footnote{Gordon's metric \cite{Gordon} was rediscovered several times, see Refs.~\cite{Quan1,Quan2,Stor}.}. To make this point as simple as possible, assume that three--dimensional space is flat (without curvature) and expands in time (Fig.~\ref{expansion}a) such that distances $\ell$ grow in time $t$ by some factor $n(t)$ as $\ell(t)=n(t)\ell_0$. Now imagine (Fig.~\ref{expansion}b) another space filled with a dielectric medium\footnote{For most dielectric materials one needs to consider not only the refractive index $n$, but the electric permittivity $\varepsilon$ and the magnetic permeability $\mu$. Both give the index, $n^2=\varepsilon\mu$, but also the impedance $Z$  of a medium, with $Z^2=\mu/\varepsilon$; in impedance--matched media and in curved space--time \cite{Plebanski} $\varepsilon=\mu=n$, which we assume throughout this paper.} of refractive index $n$ evolving in time as $n(t)$.  To electromagnetic waves --- light --- both spaces appear exactly the same. So, if attention is restricted to the quantum fluctuations of the electromagnetic field, expanding flat spaces are indistinguishable from uniform media with time--dependent refractive indices. Calculations of the zero--point energy and pressure in such media will give the electromagnetic contribution to the cosmological term. Much less is known about the quantum forces of the other fields \cite{BKMM} of the Standard Model --- experimentally, nothing at all --- but it seems reasonable to assume that their net effect follows suit. 

%%%
%%%

Note that the model of expanding flat space is a realistic approximation for the universe on cosmological scales ($>\!\!\!100 \mathrm{Mpc}$). There space is empirically known \cite{Survey} to be isotropic and homogeneous. Moreover, the relative contribution $\Omega_k$ of spatial curvature has been reduced by cosmic expansion to a small value of $|\Omega_k|<0.005$ already at the time the CMB was released \cite{CurveMeas}. Space has become nearly flat, and is expanding with uniform $n(t)$. 

Mathematically, the geometry of space and time is characterized by the metric $\mathrm{d}s$ that measures the increment of proper time along a space--time trajectory \cite{LL2}; for light $\mathrm{d}s=0$ \cite{LL2}. In particular, the space--time geometry of the expanding, spatially flat universe is described by \cite{LL2}
%%%%%%
\begin{equation}
\mathrm{d}s^2=c^2\mathrm{d}t^2-n^2\mathrm{d}\bm{r}^2
\label{metric}
\end{equation}
%%%%%%
where $c$ denotes the speed of light in the absence of gravity, or, equivalently, for zero electromagnetic susceptibility. Note that the expanding universe distinguishes a global frame \cite{Harrison}, and the metric, Eq.~(\ref{metric}), is written in the corresponding space--time coordinates. In the fictitious electromagnetic analogue of the expanding universe (Fig.~\ref{expansion}b) these are the laboratory coordinates $\bm{r}$. In cosmology \cite{Peacock,Harrison}, the $\bm{r}$--coordinates are called co--moving coordinates, because they are constant for observers staying put on geodesics \cite{LL2} parallel to each other \cite{LL2}, thus co--moving with the cosmic expansion \cite{Harrison}. As $\mathrm{d}s=0$ for $\mathrm{d}\bm{r}=0$, the coordinate time $t$ is the proper time $\mathrm{d}s/c$ experienced by co--moving observers. The metric~(\ref{metric}), with these interpretations for the coordinates, lays the scene.

\subsection{Lifshitz theory}

For calculating the cosmological constant I develop in Sec.~2 a version of Lifshitz theory \cite{Scheel,LL9}. The theory, due to Lifshitz, Dzyaloshinskii, and Pitaevskii \cite{Lifshitz,LDPrepulsion,DP,LDP}, has become the well--tested, well--established theoretical tool \cite{Rodriguez} for predicting and describing experiments with quantum forces \cite{Rodriguez,KMM,Lambrecht,Levitation,CasimirEquilibrium,Decca}. Lifshitz theory is applicable to realistic materials with dispersion and loss. It has predicted \cite{LDPrepulsion}, for example, the regime of a repulsive Casimir force that was experimentally verified in quantitative detail \cite{Levitation,CasimirEquilibrium}. Lifshitz theory has also conceptual advantages \cite{Grin1} over rivalling theories\footnote{See e.g. the Introduction of Ref.~\cite{Grin1} and references cited therein.}: it starts \cite{Rytov} from a fundamental theorem, the fluctuation--dissipation theorem \cite{Scheel}, and it involves a natural and intuitive renormalization procedure. 

Renormalization is necessary, because the energy density and pressure of quantum fluctuations seem infinite in most cases \cite{Milonni,Grin1}. This infinity must be removed to lay bare the part that does physical work, usually by comparing an arrangement of dielectric bodies at finite distances with the same bodies infinitely apart \cite{Reid}. The difference in stress on each body gives the physically meaningful force. Obviously, such a procedure applies only to calculations of the forces between bodies, but not inside them. If the entire space is filled with a medium varying in space, or time, as considered here, the medium forms a single dielectric body one cannot take apart. But here also Lifshitz theory offers a natural renormalization procedure \cite{LDP,Schwinger}. The energy density and pressure is calculated twice for each point: first assuming the actual refractive--index profile and then assuming a uniform medium equal to the local value of the index. As uniform media have no reason to exert any force inside them, the difference, if finite, gives the physically relevant energy density and pressure. Note that it is essential to remove the infinite contribution locally; a hypothetical overall infinite baseline is ruled out by experiments \cite{Levitation,CasimirEquilibrium}. 

Although Lifshitz theory gives a general prescription for calculating the energy density and pressure of quantum fluctuations in media \cite{DP}, and excellent agreement with experiment for the van der Waals and Casimir forces between bodies \cite{Rodriguez}, the quantum force inside bodies was poorly understood \cite{SimpsonSurprise}. It turned out \cite{Grin1} that the unphysical, infinite contribution depends not only on the local value of $n$, but also on its derivatives\footnote{Representing an inhomogeneous medium by infinitesimal, piece--wise homogeneous media does not give a converging Casimir force either \cite{Simpson}.}. Otherwise, the difference in Casimir stress is not finite \cite{Grin1}. This problem does not become apparent in regions of constant $n$ --- in piece--wise homogeneous media, {\it i.e.} between dielectric bodies immersed in a uniform background \cite{Levitation,CasimirEquilibrium,LDP}. For calculating quantum forces inside bodies --- in inhomogeneous media, progress has been made only recently \cite{Grin1}. This paper builds upon our work \cite{Grin1,Simpson,PXL,LeoSimpson,Grin2,Avni} on the Casimir force inside dielectrics, and upon the work of others \cite{KSW,PhilbinQ1,PhilbinQ2,Buhmann,Horsley,HorsleyPhilbin} on the quantum theory of light in media that, hopefully, may shed some light on dark energy.

One may pause here and wonder whether the renormalization does not remove the most significant contribution of the quantum vacuum to gravity. The argument goes as follows. While it is acceptable that AMO quantum forces originate from only part of the total energy density and pressure --- the renormalized part, gravity perceives everything. On the right--hand side of Einstein's equations stands the total energy--momentum tensor \cite{LL2}. In the conventional picture of Casimir forces, the total vacuum energy density $\epsilon_\mathrm{vac}$ is the infinite sum of all the zero--point energy densities of the modes involved. While only part of the sum may do mechanical work, all of it should gravitate as mass density $\epsilon_\mathrm{vac}/c^2$.

Yet Lifshitz theory offers also an alternative picture\footnote{For a visualization, see Ref.~\cite{Grin1}.} due to Schwinger \cite{SchwingerSource}. The fluctuations of the electromagnetic field originate from the sources of the field \cite{KSW,PhilbinQ1,PhilbinQ2,Buhmann,Horsley,HorsleyPhilbin,SchwingerSource}. The sources are the quantum--fluctuating charges and currents the fluctuation--dissipation theorem \cite{Scheel} requires to exist in the medium. The source fluctuations propagate as field fluctuations with the classical electromagnetic Green function as propagator \cite{KSW,PhilbinQ1,PhilbinQ2,Buhmann,Horsley,HorsleyPhilbin,SchwingerSource}. In this picture, renormalization is not the mere extraction of the mechanical energy from the infinity zero--point energy, but the removal of an artefact in the theory: the interaction of each source with itself. Here one needs to take into account the local environment, as the spurious self--interaction depends on it \cite{Grin1}. There is another caveat. Getting a finite result after renormalization is necessary, but not sufficient for obtaining the physically relevant energy density and pressure, as the renormalization may introduce artificial, finite contributions. But here also Lifshitz theory suggests physically motivated, heuristic arguments for the correct renormalization \cite{Grin1}. 

So, in Schwinger's picture \cite{SchwingerSource}, the energy and pressure of the cosmological constant is exactly proportional to the AMO vacuum energy and pressure in spatially uniform media with time--dependent refractive index.

\subsection{Objections}

One may immediately raise three objections against the chances of AMO Casimir theory in cosmology. First, transforming the time coordinate $t$ to 
%%%%%%
\begin{equation}
\tau = \int \frac{\mathrm{d}t}{n}
\label{tau}
\end{equation}
%%%%%%
transforms Eq.~(\ref{metric}) to
%%%%%%
\begin{equation}
\mathrm{d}s^2=n^2(c^2\mathrm{d}\tau^2-\mathrm{d}\bm{r}^2) \,.
\label{conformallyflat}
\end{equation}
%%%%%%
The metric has become conformally flat\footnote{A conformally flat metric may differ from the metric of flat space--time by an overall prefactor that may depend on space and time.}; $\tau$ is called conformal time. Light rays, with $\mathrm{d}s=0$, do not depend on the prefactor of the metric, and so light propagates in $\{\tau,\bm{r}\}$ coordinates like in empty, flat Minkowski space-time. As Maxwell's equations are conformally invariant\footnote{The conformal invariance of Maxwell's equations is easily seen with the help of Plebanski's interpretation of geometries as media \cite{Plebanski}. Any prefactor in the metric drops out of the constitutive equations.} this remains true for electromagnetic fields and their fluctuations. Since the renormalized vacuum energy and pressure vanishes in uniform, static media, the cosmological constant should be identically zero. 

Second, even if the AMO Casimir energy is not zero, it can only depend on derivatives of $n(t)$. On cosmological length scales where space is uniform, $n$ varies on time scales of $10^{10}$ years \cite{Peacock}, but the resulting vacuum force should dominate the dynamics of the universe. How can such a slow variation exert such a significant force? 

Third, the cosmological term in Einstein's equations \cite{LL2} appears like a fluid with positive energy density $\epsilon_\Lambda$ and negative pressure \cite{Peacock}
%%%%%%
\begin{equation}
p_\Lambda = -\epsilon_\Lambda \,,
\label{plambda}
\end{equation}
%%%%%%
whereas the vacuum pressure $p_\mathrm{vac}$ of the electromagnetic field is related to $\epsilon_\mathrm{vac}$ as
%%%%%%
\begin{equation}
p_\mathrm{vac} = \frac{1}{3}\, \epsilon_\mathrm{vac} \,,
\label{pvac}
\end{equation}
%%%%%%
    because the trace of the electromagnetic energy--momentum tensor \cite{LL2} vanishes\footnote{Here is a simple physical argument for Eq.~(\ref{pvac}). Pressure is the momentum transfer over an infinitesimal surface in infinitesimal time. For propagation with the speed of light, the momentum $P$ is related to the energy $E$ as $P=E/c$ \cite{LL2}. From this follows that the pressure is equal to the part of the energy density transported over the surface --- in one specific direction. Assuming all three directions to be equal, we arrive at Eq.~(\ref{pvac}).}. How can, for positive $\epsilon_\mathrm{vac}$, the pressure become negative and equal in magnitude to the energy density? This hypothetical, ultra--strong negative pressure should drive the expansion of the universe \cite{Peacock}. So related to this problem is the question: how can the AMO Casimir force in spatially uniform, time--dependent media become repulsive?

\subsection{Cosmological horizons}

\begin{figure}[t]
\begin{center}
\includegraphics[width=17pc]{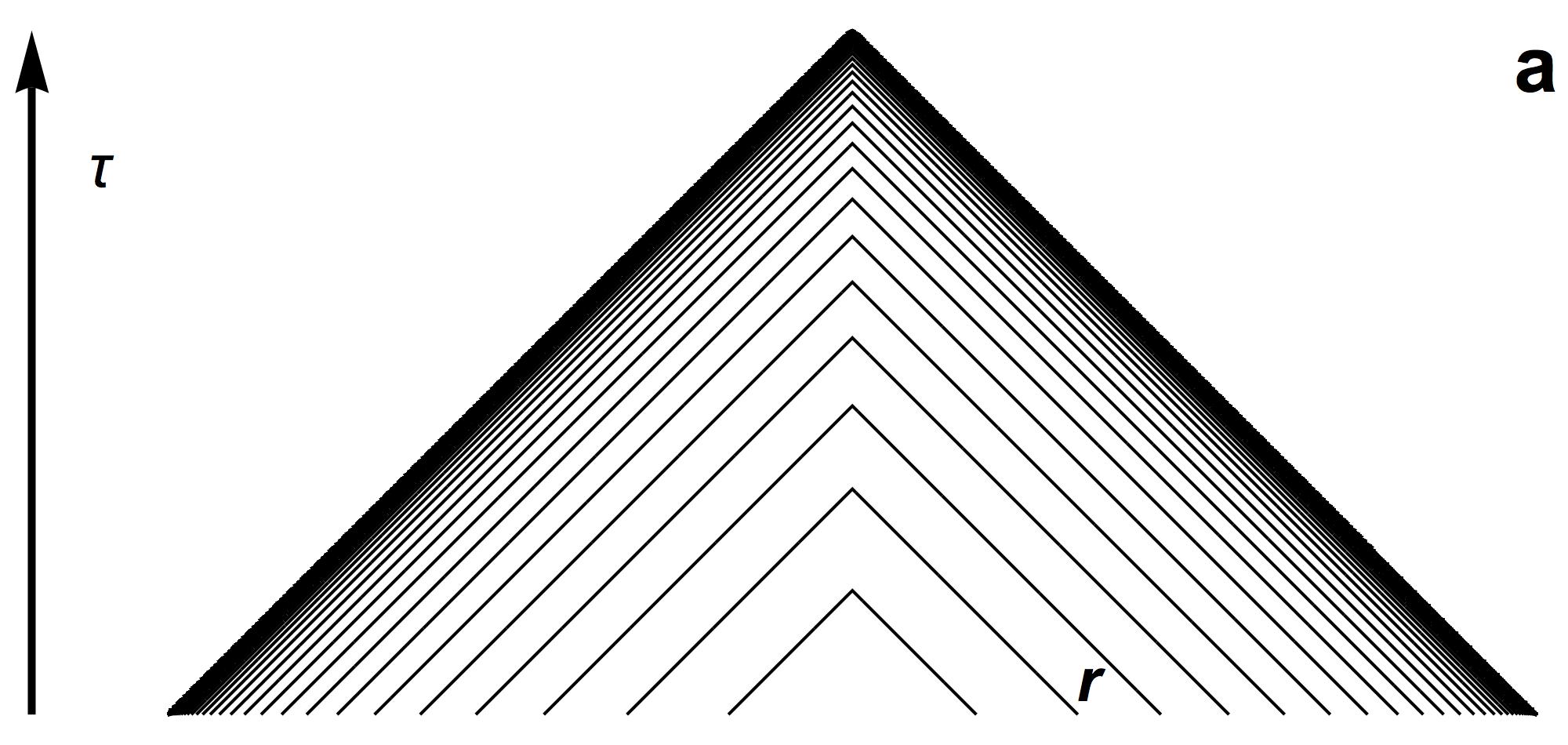}
\hspace{1pc}
\includegraphics[width=17pc]{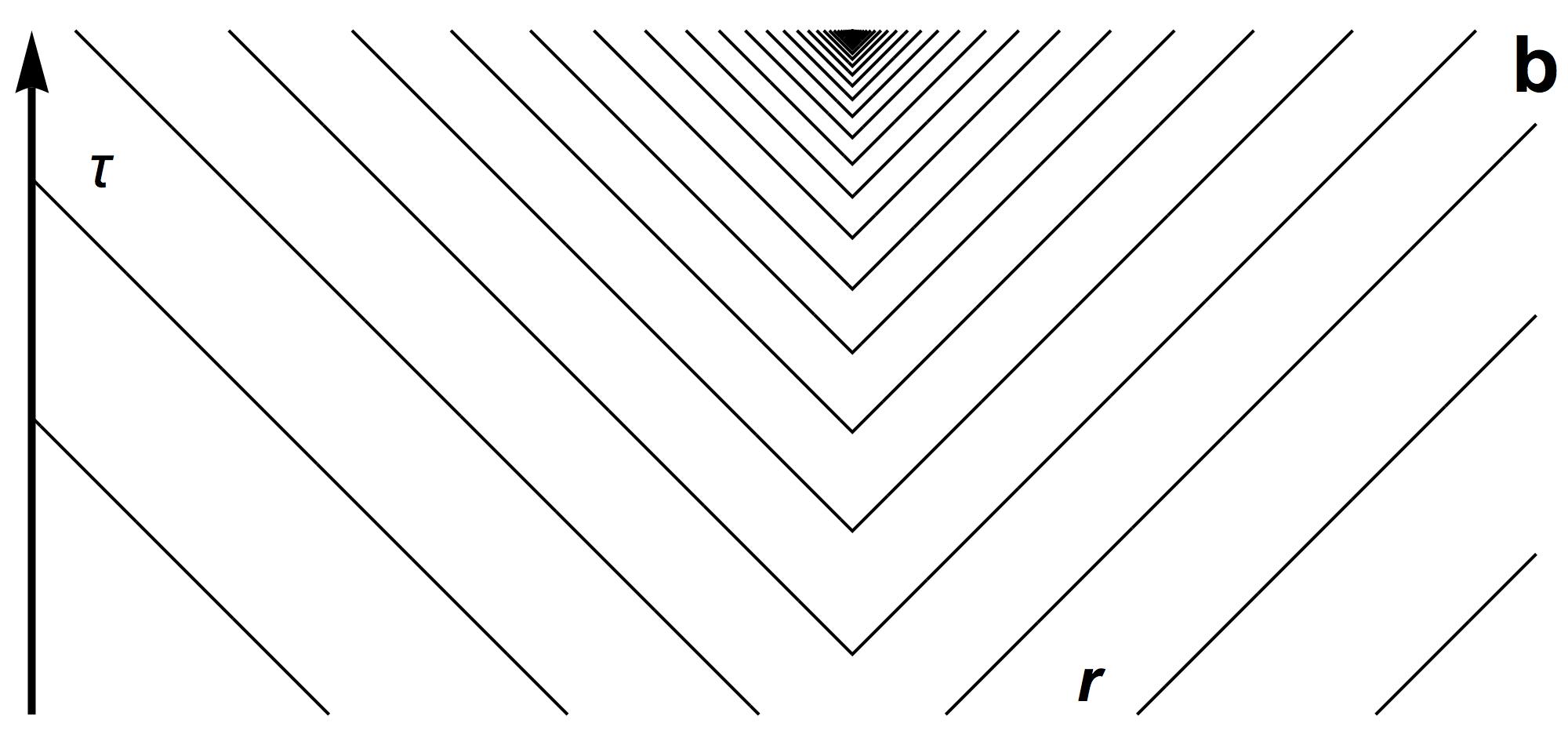}
\caption{
\small{
Cosmological horizon. The expanding universe appears like an outward moving medium around any arbitrary point in space in a coordinate system affixed to that point. The flow velocity follows Hubble's law, Eq.~(\ref{hubbleflow}), reaching the speed of light at the horizon. The figure shows typical wave fronts in conformal time $\tau$ and co--moving coordinates $\bm{r}$. {\bf a}: Phase fronts of the last waves incoming, against the Hubble flow, from the horizon, reaching the observer in the center (Fig.~\ref{expansion}a). The waves are reflected at the center and move out with phase fronts shown in {\bf b}. As outgoing waves, they propagate with the Hubble flow and so are free to cross the horizon. (Mathematical expressions for the waves are given in Appendix A.) 
}
\label{horizon}}
\end{center}
\end{figure}

Let me remove the objections of Sec.~1.3 one by one. Although one can transform the expanding universe to flat Minkowski space--time for electromagnetic fields, I will argue that cosmological horizons \cite{Harrison} remain essential (Fig.~\ref{horizon}). Since Bekenstein's \cite{Bekenstein} and Hawking's \cite{Hawking1,Hawking2} theory of black holes, horizons are known to emit thermal radiation \cite{Brout}. The transformed space--time is therefore not in a vacuum state, but in a thermal state \cite{GibbonsHawking}.

The easiest way of seeing this mathematically is by transforming the co--moving coordinates $\{\bm{r}\}$ to another set $\{\bf{r}\}$ of spatial coordinates:
%%%%%%
\begin{equation}
{\bf r} = n(t)\bm{r} \,.
\label{fix}
\end{equation}
%%%%%%
The upright $\bf{r}$--coordinates absorb the expansion factor and thus appear fixed. They are affixed to the origin, $\bf{r}=0$, the only point where they agree with the co--moving $\bm{r}$--coordinates. To see how the so--affixed coordinates experience expanding space, one transforms the metric, Eq.~(\ref{metric}), with the result:
%%%%%%
\begin{equation}
\mathrm{d}s^2=c^2\mathrm{d}t^2-(\mathrm{d}{\bf r} - H{\bf r}\, \mathrm{d}t)^2
\label{movingmetric}
\end{equation}
%%%%%%
where $H$ denotes the Hubble constant \cite{LL2}:
%%%%%%
\begin{equation}
H = \frac{\dot{n}}{n} \,.
\label{hubble}
\end{equation}
%%%%%%
Throughout this paper dots denote time derivatives. The Hubble constant is only constant for exponentially varying $n$, but the term ``constant'' is commonly used \cite{Peacock}. The transformed metric, Eq.~(\ref{movingmetric}), has an interesting physical interpretation \cite{Unruh81,Volovik,Visser,Unsch,Faccionotes}: it describes a moving fluid with radially symmetric flow speed
%%%%%%
\begin{equation}
v = H r \,.
\label{hubbleflow}
\end{equation}
%%%%%%
The expanding universe thus appears as an outward--moving fluid in the $\bf{r}$--coordinates affixed to the co--moving observer at the origin. As one can shift the origin to any other point, this is also true for all other observers co--moving with the cosmic background. For each and everyone of them, the universe flows away with radial velocity $v$. Equation (\ref{hubbleflow}) describes one of Hubble's laws \cite{LL2}: the expansion velocity grows linearly with growing distance. At some radius, the flow velocity $v$ reaches the speed of light $c$. No classical communication from a sphere of greater radius is possible; the radius where $v=c$ defines a cosmological horizon (Fig.~\ref{horizon}) \cite{Harrison}.

%%%
%%%

Each co--moving point in space is surrounded by a cosmological horizon, separating the world into an inside and an outside. The inside sphere depends on the co--moving point and encloses different spatial regions for different points. The quantum vacuum, however, is universal; the vacuum cannot possibly adjust itself to all the conflicting horizons, and so it must remain indivisible. The vacuum bridges the dividing horizons, entangling the field in the inside with the field outside of each horizon in Einstein--Podolsky--Rosen (EPR) states \cite{EPR,LeoBook,Classical}. To an inside observer, the pure EPR state appears as a statistical mixture with maximal entropy \cite{LeoBook,Barnett}: a thermal state. In moving fluids \cite{Unruh81,Volovik,Visser,Unsch,Faccionotes}, the temperature is given by the velocity gradient at the horizon. For cosmological horizons, I obtain the Gibbons--Hawking temperature \cite{GibbonsHawking} (generalized here beyond de Sitter space \cite{deSitter}):
%%%%%%
\begin{equation}
k_\mathrm{B} T = \frac{\hbar H}{2\pi}
\label{gh}
\end{equation}
%%%%%%
where $k_\mathrm{B}$ denotes Boltzmann's constant and $\hbar$ the reduced Planck constant. I deduced the Gibbons--Hawking temperature with the help of the coordinates affixed to a co--moving observer, Eq.~(\ref{fix}), but the quantum radiation of cosmological horizons does not disappear in other coordinate systems, only the radiation temperature, Eq.~(\ref{gh}), changes when the measure of time is changed. In conformal time $\tau$ one gets
%%%%%%
\begin{equation}
k_\mathrm{B} \Theta = \frac{\hbar\dot{n}}{2\pi}
\label{conftemp}
\end{equation}
%%%%%%
as the ratio of frequency and temperature must remain invariant, with the frequency changing by a factor of $n$ according to Eq.~(\ref{tau}) and $H$ being given by Eq.~(\ref{hubble}). 

\subsection{Equivalence principle}

In conformal coordinates $\{\tau,\bm{r}\}$ space--time appears flat to electromagnetic waves, but it is not empty: it is filled with thermal radiation. Yet for a Hubble constant in the order of $1$ per $10^{10}$ years ($3\times 10^{-18}\mathrm{Hz}$), the Gibbons--Hawking temperature lies around $2\times 10^{-29}\mathrm{K}$. How can such a tiny temperature compete with the $2.7\mathrm{K}$ of the CMB \cite{Peacock}? From Lifshitz theory in ordinary media one will certainly not expect a significant figure for the Casimir force in media varying on the time scale of $10^{10}$ years. What can be different between ordinary media and space--time? Did I not argue that they are the same? 

The equivalence principle makes all the difference. According to the equivalence principle, the space--time geometry applies to everything equally, not only to the electromagnetic field, and for electromagnetic waves not only to a small range of frequencies --- in contrast to ordinary media. The response of ordinary media varies with frequency and vanishes for frequencies beyond the atomic scale, whereas space--time should act equally on the entire spectrum --- up to the Planck scale. Close to the Planck scale space--time is expected to become dispersive \cite{Jacobson}, violating the equivalence principle. We found in our previous work \cite{Grin1} that the renormalization of the Casimir force relies critically on the fact that the response of physical media drops sufficiently fast with frequency. Without this, the Casimir stress in planar, inhomogeneous media would become infinite in general. As this attempted infinity appears in renormalization, it is not influenced by an ordinary thermal background like the CMB. But in renormalization, the vast spectrum seen by gravity may turn the tiny Gibbons--Hawking temperature of the expanding universe into a quantity deciding its fate. 

Let me estimate what it takes for the vacuum energy to have a significant influence on the cosmic dynamics. Einstein's equations in a homogeneous and isotropic space are reduced to the two Friedman equations \cite{LL2,Friedman1,Friedman2}. In flat space, the first Friedman equation relates the total energy density $\epsilon$ to the Hubble constant \cite{LL2}:
%%%%%%
\begin{equation}
H^2 = \frac{8\pi G}{3c^2}\,\epsilon
\label{f1}
\end{equation}
%%%%%%
where $G$ denotes Newton's gravitational constant. Suppose that $\epsilon$ is essentially given by the vacuum energy density $\epsilon_\mathrm{vac}$ (the general case is considered later in Sec.~4). Being ultimately a quantum energy, whether directly or indirectly via the Gibbons--Hawking temperature, $\epsilon_\mathrm{vac}$ must be proportional to $\hbar$. In order to influence the dynamics following Friedman's Eq.~(\ref{f1}), $\epsilon_\mathrm{vac}$ should scale as $H^2$. Having the physical dimensions of an energy density implies that $\epsilon_\mathrm{vac}$ should be equal to $\hbar H^2/c$ times a dimensionless constant divided by the square of a length. If I set this length to the order of the Planck length with
%%%%%%
\begin{equation}
\ell_\mathrm{p} = \sqrt{\frac{\hbar G}{c^3}} 
\label{planck}
\end{equation}
%%%%%%
Friedman's Eq.~(\ref{f1}) is satisfied. So, if the renormalized energy density of the quantum vacuum diverges with an inverse length squared in an ideal space--time \cite{LL2} honoring the equivalence principle indefinitely, the vacuum energy becomes cosmologically relevant for a realistic space--time \cite{Jacobson} respecting the equivalence principle only up to the Planck scale.

In planar media, where $n$ varies only in one direction in space and is otherwise constant, the Casimir stress diverges logarithmically with the frequency cut--off \cite{Grin1}. Media with $n$ constant in space but varying in time are different though, due to the existence of horizons and, as will be seen in Sec.~2, causality. 

\subsection{Trace anomaly}

The second Friedman equation \cite{Friedman1,Friedman2} follows from the conservation of energy and momentum\footnote{Conservation of energy and momentum means that the covariant divergence of the energy--momentum tensor vanishes \cite{LL2}:
$ 0=D_\mu T^{\mu\nu} = \frac{1}{\sqrt{-g}}\partial_\mu \left(\sqrt{-g}T^\mu_\nu\right)-\frac{1}{2}\left(\partial_\nu g_{\alpha\beta}\right) T^{\alpha\beta} $
here with energy momentum tensor $T^\mu_\nu=\mathrm{diag}(\epsilon,p,p,p)$ and metric tensor $g_{\alpha\beta}=\mathrm{diag}(1,-n^2,-n^2,-n^2)$. From this follows Eq.~(\ref{f2}). For a purely thermodynamic derivation of the second Friedman equation see Ref.~\cite{LL2}.}. In a spatially isotropic and homogeneous universe, all energy and matter must move on average with the expanding cosmic background. For a fluid of energy density $\epsilon$ and pressure $p$ following adiabatically the expansion of the universe, entropy must be conserved \cite{LL5}. The conservation of entropy is part of relativistic fluid mechanics \cite{LL6}; in the absence of any net transport relative to the cosmic background it becomes the only non--trivial aspect of energy--momentum conservation \cite{LL6}. One obtains from thermodynamics the second Friedman equation \cite{LL2}:
%%%%%%
\begin{equation}
\dot{\epsilon} = -3(\epsilon+p) H \,.
\label{f2}
\end{equation}
%%%%%%
The pressure $p_\mathrm{vac}$ of the electromagnetic vacuum is given in terms of the energy density $\epsilon_\mathrm{vac}$ by the equation of state, here Eq.~(\ref{pvac}). The energy density must be a function of derivatives of the refractive index, 
%%%%%%
\begin{equation}
\epsilon_\mathrm{vac} = f(n,\dot{n}, ...) \,,
\label{ef}
\end{equation}
%%%%%%
as the renormalized vacuum energy of a uniform medium vanishes. The dependence on derivatives implies, however, that the quantum vacuum is not adiabatic. Friedman's Eq.~(\ref{f2}), obtained from adiabaticity \cite{LL2}, combined with Eqs.~(\ref{pvac}), (\ref{hubble}) and (\ref{ef}), would give:
%%%%%%
\begin{equation}
\dot{\epsilon}_\mathrm{vac} = (\partial_n f) \dot{n} + (\partial_{\dot{n}} f) \ddot{n} + ... = -4\frac{\dot{n}}{n}\,.
\label{nonadiabatic}
\end{equation}
%%%%%%
This equation needs to be satisfied for all possible $n(t)$, for otherwise it would define a differential equation for $n(t)$ in conflict with the first Friedman equation, Eq.~(\ref{f1}). Equation (\ref{nonadiabatic}) is satisfied for all $n(t)$ only when all the terms in front of higher time derivatives of $n$ vanish, from the highest to the first derivative, and when $\partial_n f = -4/n$, so for $f\propto n^{-4}$, which gives the standard equation of state for radiation in the expanding universe \cite{LL2}. There is no room for maneuver to include derivatives. 

The quantum vacuum violates adiabaticity and hence energy--momentum conservation. Wald \cite{Wald} discovered the root of the problem: the lack of reciprocity in the renormalization procedure. In Schwinger's \cite{SchwingerSource} picture of Lifshitz theory \cite{Grin1}, the problem takes the following form. Each source is split into an emitter and a receiver infinitesimally apart. The self--interaction of the source depends on the local environment of the emitter, but not on the environment of the receiver. Emitter and receiver are not reciprocal, which --- using non--technical language --- causes an imbalance in recoil that appears as an additional energy and pressure. In technical terms, the lack of reciprocity in renormalization causes a trace anomaly \cite{Wald}. 

Suppose, for simplicity, that $\epsilon$ and $p$ are solely given by the quantum vacuum, including the recoil imbalance (trace anomaly). If I write
%%%%%%
\begin{equation}
\epsilon = \epsilon_\mathrm{vac} + \epsilon_\Lambda \,,\quad
p = p_\mathrm{vac} -  \epsilon_\Lambda
\label{ansatz}
\end{equation}
%%%%%%
the right--hand side of Friedman's Eq.~(\ref{f2}) is not affected, but the left--hand side gets an additional term $\dot{\epsilon}_\Lambda$ taking care of energy--momentum conservation: $\epsilon_\Lambda$ is the missing recoil energy density. The notation is suggestive. In Eq.~(\ref{ansatz}) the energy density $\epsilon_\Lambda$ is accompanied by a pressure of equal magnitude but opposite sign, exactly like in Eq.~(\ref{plambda}) the pressure $p_\Lambda$ of the cosmological constant. Let me represent $\epsilon_\mathrm{vac}$ by
%%%%%%
\begin{equation}
\frac{4\pi G}{3c^2}\,\epsilon_\mathrm{vac} = -\alpha_\Lambda \Delta 
\label{delta}
\end{equation}
%%%%%%
where $\alpha_\Lambda$ is a dimensionless constant and $\Delta$ a function of $n$ and its derivatives, with the dimension of a frequency squared. Differentiating the first Friedman equation, Eq.~(\ref{f1}), with respect to time, and applying the second Friedman equation, Eq.~(\ref{f2}), I obtain an equation of motion for the cosmic expansion driven by the quantum vacuum:
%%%%%%
\begin{equation}
\dot{H} = 4\alpha_\Lambda \Delta \,,
\label{dynamics}
\end{equation}
%%%%%%
which establishes a differential equation for the expansion factor $n(t)$, as both $H$ and $\Delta$ depend on $n$ and its derivatives. Furthermore, I get directly from Friedman's Eq.~(\ref{f1}) the energy $\epsilon_\Lambda$ and hence the cosmological constant \cite{Peacock}
%%%%%%
\begin{equation}
\Lambda = \frac{8\pi G}{c^4} \,\epsilon_\Lambda = \frac{3}{c^2} \left(H^2 + \frac{2}{3} \alpha_\Lambda \Delta\right) \,.
\label{lambda}
\end{equation}
%%%%%%
Given a solution of the dynamics, Eq.~(\ref{dynamics}), $\Lambda$ is determined. As argued in Sec.~1.5, the quantum vacuum dominates the dynamics if the energy density $\epsilon_\mathrm{vac}$ diverges with the inverse square of a length set to the order of the Planck length, Eq.~(\ref{planck}). In this case, $\alpha_\Lambda$ lies in the order of unity. 

\subsection{Results}

In Sec.~2 I calculate the renormalized electromagnetic vacuum energy density in spatially uniform, time dependent media. I find
%%%%%%
\begin{equation}
\Delta = \partial_t^3 \frac{1}{H} + H \partial_t^2 \frac{1}{H}
\label{result}
\end{equation}
%%%%%%
where $\partial_t$ abbreviates the time derivative $\partial/\partial t$. In Sec.~2 I also express $\alpha_\Lambda$ in terms of the cut--off for the electromagnetic contribution to the vacuum energy. For the other fields of the Standard Model, the quantum noise will be linear, even for non--Abelian fields with non--linear field equations \cite{Weinberg2}. Massive fields are known \cite{BKMM} to modify the standard Casimir force, because their amplitudes decay in propagation, reducing the reflection amplitude the force relies on \cite{Scheel}, but the renormalization of Sec.~2 is local and hence should not be affected. Therefore it seems reasonable that the principal structure of the result, Eqs.~(\ref{delta}) and (\ref{result}), extends beyond quantum electromagnetism, but with a different $\alpha_\Lambda$ taking into account the sum of the contributions of the other fields of the Standard Model. 

As expected, $\Delta$ and hence $\epsilon_\mathrm{vac}$ depends on derivatives of the refractive index $n$, up to forth order, according to the definition (\ref{hubble}) of the Hubble constant. Equation~(\ref{dynamics}) shows that, in the absence of any other energy and matter, one can multiply $n$ by any constant scale factor and have the same dynamics. So the evolution of the spatially flat, empty universe is independent of its size, which is natural, as there are no other length scales involved\footnote{The inverse Planck length squared in the cosmological energy density compensates for the natural constants in the Friedman equations \cite{LL2} such that the equation of motion in flat space, Eq.~(\ref{dynamics}), does not depend on a length scale.}. The situation is different in curved space, but Eqs.~(\ref{delta}) and (\ref{result}) for the vacuum energy remain the same, as I show in Sec.~3. 

Equation (\ref{result}) implies that the cosmological constant of Eq.~(\ref{lambda}) is no longer constant, except in de Sitter space \cite{Harrison,deSitter} where $H=\mathrm{const}$. Here the universe is expanding exponentially, as Eq.~(\ref{hubble}) has the solution $n=n_0\exp(-Ht)$. In general, the cosmic expansion creates the vacuum energy that, in turn, corrects the expansion. In exponentially expanding de Sitter space, no correction is required; de Sitter space is a consistent solution of Einstein's equations of gravity and quantum field theory. 

How does the interplay between the cosmic expansion and the quantum vacuum react to small perturbations of the exponential expansion in flat space? Consider
%%%%%%
\begin{equation}
\frac{1}{H} = \frac{1}{H_0} + \xi(t)
\label{correction}
\end{equation}
%%%%%%
with constant $H_0$ and $\xi$ small in comparison with $H_0^{-1}$. In Eq.~(\ref{dynamics}) write $\dot{H}$ as $-H^2\partial_t H^{-1}$, linearize in $\xi$ with Eq.~(\ref{result}) for $\Delta$, and integrate the result in time, absorbing the integration constant in the choice of $H_0$. One arrives at the equation of a damped harmonic oscillator:
%%%%%%
\begin{equation}
\ddot{\xi} + H_0\, \dot{\xi} +\frac{H_0^2}{4\alpha_\Lambda}\, \xi = 0 
\label{oscillator}
\end{equation}
%%%%%%
with damping rate $H_0/2$ of the amplitude $\xi$. Small perturbations of de Sitter expansion are damped and so corrected for by the quantum vacuum (assuming of course that the vacuum dominates the cosmic expansion).

Lifshitz theory thus predicts that the cosmological constant $\Lambda$ is not constant, but a dynamical quantity similar to quintessence \cite{Caldwell}. Perturbations of $\Lambda$ should last in the order of the Hubble constant. There are indeed some indications from astronomical observations that $\Lambda$ has varied. The directly observed Hubble constant from galactic cepheids \cite{Super} differs from the calculated $H$ using the $\Lambda$ from CMB measurements \cite{CMBPlanck} by $6\%$, which suggests that the cosmological constant at the time of the formation of the cosmic background radiation was different than the present $\Lambda$. 

\subsection{Methodology}

\begin{figure}[t]
\begin{center}
\includegraphics[width=20.5pc]{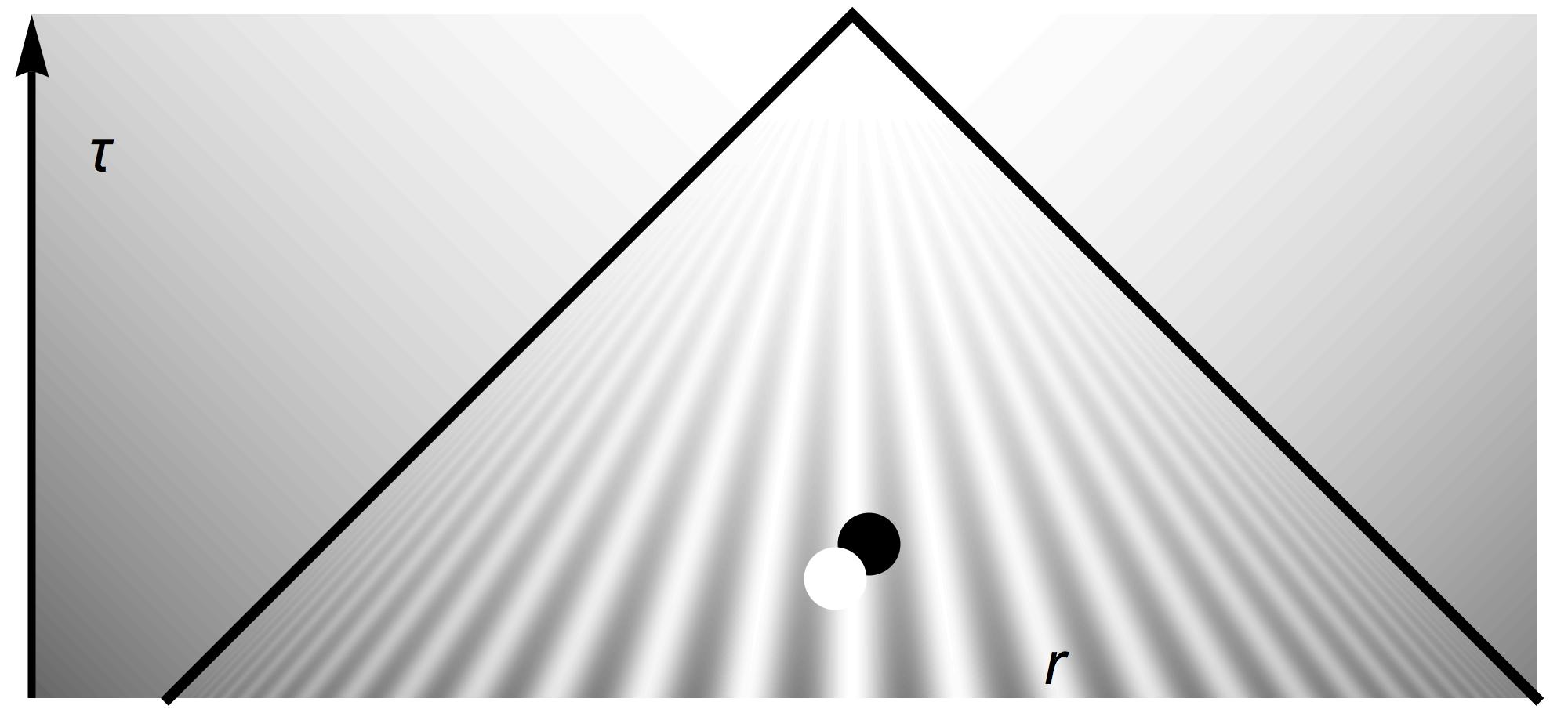}
\caption{
\small{
Methods. Points in space and time in conformal coordinates $\{\tau,\bm{r}\}$ are split into an emitter (white dot) and a receiver (black dot) infinitesimally close to each other. Causality requires that emission precedes reception. The cosmological horizon (black lines) generates thermal radiation influencing the propagation of field fluctuations between emitter and receiver. The radiation consists of superpositions of incoming and outgoing waves (Fig.~\ref{horizon}). The intensity pattern (level of brightness) of one of such waves is shown (mathematical details in Appendix A). 
}
\label{method}}
\end{center}
\end{figure}

Figure \ref{method} illustrates the physical assumptions behind the mathematical method applied in the calculations of Secs.~2 and 3. First, in order to identify the self--interaction of each source to be subtracted in renormalization, one imagines each point in space and time as being split into two: an emitter and a receiver. This is known as the point--splitting method. 

Second, in the self--interaction the emission depends on the local environment of the emitter. In our previous work \cite{Grin1} we found that, in planar media, one should take into account the local value and the derivatives of the refractive index $n$ up to second order. Here I assume the same for time--dependent media: in the self--interaction $n(t)$ is set to a quadratic function around the time of emission. 

Third, there is an important difference between point--splitting in space and event--splitting in time: causality. While we can go back and forth in space, we cannot do so in time; the time of emission must precede the time of reception. Section~2 shows that causality, combined with the second--order expansion of the local environment, causes a subtle discontinuity in the renormalization that produces a divergence of the energy density with an inverse length squared. As argued in Sec.~1.5, this singularity allows the quantum vacuum to influence the dynamics of the universe. It naturally comes from causality.

Fourth, another important difference to the planar case \cite{Grin1} is the existence and radiation of cosmological horizons. I argued in Sec.~1.4 that horizons are necessary for the renormalized vacuum energy of uniform space to be different from zero, and this also follows naturally from the theory of Sec.~2 without making additional assumptions. 

Fifth, my starting point is the same as in Lifshitz' original paper \cite{Lifshitz}: the fluctuation--dissipation theorem \cite{Scheel}. In the context of quantum forces \cite{Forces} the theorem relates the fluctuations of the electromagnetic field to the dissipation during propagation, depending on temperature. In order to define a temperature, one needs a Hamiltonian \cite{LL5}. In media with time--dependent refractive index, a Hamiltonian exists only for the free propagation in conformal time, Eq.~(\ref{tau}). The horizon temperature however, Eq.~(\ref{conftemp}), depends on time. In the fluctuation--dissipation theorem \cite{Scheel} one needs to identify a definite temperature and hence a definite time, even in the limit of the infinitesimally close emission and reception time taken in the point--splitting method. It appears natural to assume that the temperature should be taken at the conformal time exactly between emission and reception (Fig.~\ref{method}).

This paper does make assumptions and extrapolates Lifshitz theory vastly beyond the experimentally tested validity range of AMO Casimir physics \cite{Rodriguez,KMM,Lambrecht,Levitation,CasimirEquilibrium,Decca}, but the assumptions are grounded in proven physical principles, are not specific to the cosmological constant, and some if not all are experimentally testable, if not directly then in laboratory analogues \cite{Visser,FedFish}. For example, the concepts from the Casimir theory of planar media \cite{Grin1} have physical consequences in the aggregation in liquids \cite{Grin2} where they can be tested. The theory of this paper should and can be confronted with empirical data, as it does make quantitative predictions. The paper is conservative in the physics --- no new fields are introduced to explain the cosmological constant, but rather new concepts in fields as old as quantum electromagnetism. 

%%%
%%%

\section{Calculation}

This section substantiates the arguments of Sec.~1 with calculations, starting from quantum electromagnetism in media \cite{Leo}. Although I apply Schwinger's source theory \cite{SchwingerSource} and the related quantum theories of light in dispersive and dissipative media \cite{KSW,PhilbinQ1,PhilbinQ2,Buhmann,Horsley,HorsleyPhilbin} as the guiding principle for renormalization, the actual calculations are performed in dispersionless materials where more conventional theories are at hand. As argued in Sec.~1.5, neglecting dispersion is justified up to the Planck scale \cite{Jacobson} according to the equivalence principle. 

\subsection{Quantum electromagnetism}

Consider the electromagnetic quantum field in media \cite{Leo} with spatially uniform, but time--dependent refractive index $n(t)$, and equal electric and magnetic response. For quantum electromagnetism in media \cite{Leo} it is advantageous to impose the Coulomb gauge  on the operator $\hat{\bm{A}}$ of the electromagnetic vector potential:
%%%%%%
\begin{equation}
\nabla \cdot n \hat{\bm{A}} = 0 \,,
\label{coulomb}
\end{equation}
%%%%%%
which, for spatially uniform media, implies $\nabla \cdot \hat{\bm{A}} = 0$. The operators of the electric field strength $\hat{\bm{E}}$, the dielectric displacement $\hat{\bm{D}}$, the magnetic induction $\hat{\bm{B}}$ and the magnetic field strength $\hat{\bm{H}}$ are given in terms of the vector potential $\hat{\bm{A}}$ and in SI units \cite{Jackson}: 
%%%%%%
\begin{equation}
\hat{\bm{E}} = - \partial_t \hat{\bm{A}} \,,\quad \hat{\bm{D}} = \varepsilon_0 n \hat{\bm{E}} \,,\quad
\hat{\bm{B}} = \nabla\times\hat{\bm{A}} \,,\quad \hat{\bm{H}} = \frac{\varepsilon_0 c^2}{n}\hat{\bm{B}}
\label{fields}
\end{equation}
%%%%%%
where $\varepsilon_0$ denotes the electric permittivity of the vacuum \cite{Jackson}. From the electromagnetic energy density in dispersionless media \cite{Jackson} as Hamilton density
%%%%%%
\begin{equation}
\hat{\mathcal{H}} = \frac{1}{2} \left( \hat{\bm{E}} \cdot \hat{\bm{D}} + \hat{\bm{B}} \cdot \hat{\bm{H}} \right)
\end{equation}
%%%%%%
follows \cite{Leo} Maxwell's equations if one requires the canonical commutation relation, 
%%%%%%
\begin{equation}
\big[ \hat{\bm{D}}(\bm{r}_1,t), \hat{\bm{A}}(\bm{r}_0,t)\big] = \mathrm{i}\hbar\, \delta^\mathrm{T}(\bm{r}_1-\bm{r}_0)
\label{commutator}
\end{equation}
%%%%%%
where $\delta^\mathrm{T}(\bm{r})$ denotes the transversal delta function \cite{Leo,MandelWolf}. Note that the results of this paper are gauge--invariant, as they only depend on the energy density and hence on the fields of Eq.~(\ref{fields}) that are themselves gauge--invariant. Equations (\ref{coulomb}-\ref{commutator}) contain all ingredients needed from quantum electromagnetism. 

\subsection{Point splitting}

In the point--splitting method (Fig.~\ref{method}) applied in the calculations of this paper, one considers first the correlation function for the electromagnetic energy density in the vacuum state
%%%%%%
\begin{equation}
u_\mathrm{vac} \equiv \frac{1}{4} \langle \hat{\bm{E}}_1 \cdot \hat{\bm{D}}_0 + \hat{\bm{E}}_0 \cdot \hat{\bm{D}}_1 + \hat{\bm{B}}_1 \cdot \hat{\bm{H}}_0 + \hat{\bm{B}}_0 \cdot \hat{\bm{H}}_1 \rangle
\label{correlation}
\end{equation}
%%%%%%
with indices referring to the space--time points $(t_0,\bm{r}_0)$ and $(t_1,\bm{r}_1)$, renormalizes, and then takes the limits $t_1 \rightarrow t_0$ and $\bm{r}_1 \rightarrow \bm{r}_0$. Note that the order of operators is not important, as $\hat{\bm{E}}$ and $\hat{\bm{D}}$ are proportional to each other in Eq.~(\ref{fields}) and hence commute, and so do $\hat{\bm{B}}$ and $\hat{\bm{H}}$. However, the order of limits is important and encodes some of the physics beyond the dispersionless model. Taking the limit $t_1 \rightarrow t_0$ first means assuming the absence of dispersion for all time scales, including the Planck scale, whereas the limit $\bm{r}_1 \rightarrow \bm{r}_0$ taken first implies no spatial dispersion. Ordinary dielectric media are mostly dispersive in the time domain, as the dielectric displacement responds locally and with characteristic resonances and delay times \cite{LL8}. It turns out in Sec.~2.10 that in time--dependent uniform media the renormalized vacuum energy density diverges with an inverse length squared only in the absence of temporal dispersion. As argued in Sec.~1.5, such a divergence is required for the quantum vacuum to create the cosmological constant, which would therefore indicate that space--time is spatially dispersive \cite{LL8} on the Planck scale, as in metals, electrolytes and plasmas, and also as in acoustic analogues of gravity \cite{Unruh81,Volovik,Visser,Unsch,Faccionotes}. 

Note that the cosmological energy density $\epsilon_\mathrm{vac}$ of the quantum vacuum is the $T^0_0$ component of the energy--momentum tensor \cite{LL2}, whereas $u_\mathrm{vac} = \sqrt{-g}\,T^0_0$ \cite{LL2} with $g$ being the determinant of the metric tensor \cite{LL2}. For the metric described in Eq.~(\ref{metric}) $g=-n^6$ and hence
%%%%%%
\begin{equation}
\epsilon_\mathrm{vac} = \frac{1}{n^3} \lim_{t_1\rightarrow t_0} \left(u_\mathrm{vac}-u^0_\mathrm{vac}\right)
\label{epsvac}
\end{equation}
%%%%%%
where $u^0_\mathrm{vac}$ denotes the correlation function of Eq.~(\ref{correlation}) evaluated with $n$ being replaced by a quadratic polynomial around the time of emission (Sec.~2.7).

\subsection{Reduction to scalar field}

The electromagnetic field carries two polarizations that, in uniform media, are completely independent and equal to each other. One can therefore replace the vector potential $\hat{\bm{A}}$ in the correlation function $\langle \hat{\bm{A}}_1 \cdot \hat{\bm{A}}_0\rangle$ by two equal scalar fields $\hat{A}$ as
%%%%%%
\begin{equation}
\langle \hat{\bm{A}}_1 \cdot \hat{\bm{A}}_0\rangle = 2 \langle \hat{A}_1 \hat{A}_0 \rangle \,.
\end{equation}
%%%%%%
Furthermore, as 
%%%%%%
\begin{equation}
\big(\nabla_1 \times \hat{\bm{A}}_1\big) \cdot \big(\nabla_0 \times \hat{\bm{A}}_0\big) = (\nabla_1\cdot\nabla_0) (\hat{\bm{A}}_1\cdot\hat{\bm{A}}_0)
\end{equation}
%%%%%%
one obtains from Eqs.~(\ref{fields}) and (\ref{correlation}):
%%%%%%
\begin{equation}
u_\mathrm{vac} = \frac{\varepsilon_0 n }{2} \left(\partial_{t_0} \partial_{t_1} + \frac{c^2}{n^2} \nabla_1\cdot\nabla_0\right)
\langle \hat{A}_1 \hat{A}_0 + \hat{A}_0 \hat{A}_1 \rangle \,.
\label{uvac0}
\end{equation}
%%%%%%
As discussed in Sec.~1.8, $u_\mathrm{vac}$ needs to be described in conformal times $\tau_1$ and $\tau_0$ for application in the fluctuation--dissipation theorem.  Let me represent $\tau_1$ and $\tau_0$ as 
%%%%%%
\begin{equation}
\tau_1 = \tau +\frac{\sigma}{2} \,,\quad \tau_0 = \tau - \frac{\sigma}{2}
\label{tau10}
\end{equation}
%%%%%%
and introduce the field correlation 
%%%%%%
\begin{equation}
K \equiv \frac{\varepsilon_0 c}{2\hbar} \, \langle \hat{A}_1(\tau_1)\,\hat{A}_0(\tau_0) + \hat{A}_0(\tau_0)\,  \hat{A}_1(\tau_1) \rangle
\label{k}
\end{equation}
%%%%%%
with indices referring to the spatial coordinates $\bm{r}_1$ and $\bm{r}_0$. With these definitions, Eqs.~(\ref{tau}) and (\ref{uvac0}) give
%%%%%%
\begin{equation}
u_\mathrm{vac} = \frac{\hbar}{c n} \, \left(\frac{1}{4}\,\partial_\tau^2 - \partial_\sigma^2 + c^2\, \nabla_1\cdot\nabla_0\right) K \,,
\label{uvac}
\end{equation}
%%%%%%
which completes the preparations for the fluctuation--dissipation theorem. 

\subsection{Fluctuation and dissipation}

The field fluctuations of the quantum vacuum give rise to the field correlations of Eq.~(\ref{k}) that generate the electromagnetic energy density according to Eq.~(\ref{uvac}). Consider now
%%%%%%
\begin{equation}
\Gamma \equiv \frac{\varepsilon_0 c}{2\mathrm{i}\hbar} \, \langle \big[\hat{A}_1(\tau_1), \hat{A}_0(\tau_0)\big] \rangle = \frac{1}{2c} \left( G_+ - G_-\right)
\label{gamma}
\end{equation}
%%%%%%
where the $G_\pm$ denote the retarded (+) and the advanced (-) Green functions \cite{BD}:
%%%%%%
\begin{equation}
G_\pm = \pm \frac{\varepsilon_0 c^2}{\mathrm{i}\hbar} \,\Theta\left(\pm(\tau_1-\tau_0)\right) \, \langle \big[\hat{A}_1(\tau_1), \hat{A}_0(\tau_0)\big] \rangle 
\end{equation}
%%%%%%
and $\Theta(\tau)$ the Heaviside step function. Indeed, $G_\pm$ satisfies the homogeneous wave equation in $(\tau_1,\bm{r}_1)$ coordinates for $\tau_1\neq\tau_0$ as $\hat{A_1}$ does, and one finds from the canonical commutation relation, Eq.~(\ref{commutator}), and Eqs.~(\ref{tau}) and (\ref{fields}), that $\partial_{\tau_1}^2 G_\pm \sim -c^2\, \delta(\tau_1-\tau_0)\,\delta(\bm{r}_1-\bm{r}_0)$ for $\tau_1\sim\tau_0$, which implies that $G_\pm$ obeys the defining equation of a classical electromagnetic Green function:
%%%%%%
\begin{equation}
\left(\nabla_1^2-\frac{1}{c^2}\partial_{\tau_1}^2\right) G_\pm = \delta(\tau_1-\tau_0)\,\delta(\bm{r}_1-\bm{r}_0) \,.
\label{greeneq}
\end{equation}
%%%%%%
The solution of Eq.~(\ref{greeneq}) is well--known \cite{Jackson}:
%%%%%%
\begin{equation}
G_\pm = - \frac{1}{4\pi c\rho} \, \delta (\sigma \mp \rho)  
\label{green0}
\end{equation}
%%%%%%
with $\sigma=\tau_1-\tau_0$ according to definition (\ref{tau10}) and 
%%%%%%
\begin{equation}
\rho=\frac{r}{c} \,,\quad r = \left| \bm{r}_1-\bm{r}_0 \right| \,.
\label{r}
\end{equation}
%%%%%%
The Green function $G_\pm$ thus represents a flash of light emitted (+) or absorbed (-) at point $\bm{r}_0$. The difference $2c\Gamma$ between emission and absorption, Eq.~(\ref{gamma}), describes dissipation. 

For relating the dissipation $\Gamma$ to the fluctuation $K$, define for complex time $z$ the analytic function\footnote{For the total correlation, the function $f(z)$ defined in Eq.~(\ref{fdef}) is analytic, because the density matrix of Eq.~(\ref{thermal}) depends only on $\tau$, but not on $z$, and $z$ enters $f$ through the free--field evolution of Eq.~(\ref{free}). For the correlation of the self--interaction, see the footnote in Sec.~2.8.}
%%%%%%
\begin{equation}
f(z) \equiv \frac{\varepsilon_0 c}{\hbar} \, \langle \hat{A}_1(\tau+\frac{z}{2})\, \hat{A}_0(\tau-\frac{z}{2}) \rangle \,.
\label{fdef}
\end{equation}
%%%%%%
As the electromagnetic field is real for real times, the operators $\hat{A}$ are Hermitian for real $z=\sigma$, and hence
%%%%%%
\begin{equation}
K = \mathrm{Re} f(\sigma) \,,\quad \Gamma = \mathrm{Im} f(\sigma) 
\label{rh}
\end{equation}
%%%%%%
according to definitions (\ref{k}) and (\ref{gamma}). The fluctuation $K$ and the dissipation $\Gamma$ are thus the real and imaginary part of an analytic function on the real axis: the fluctuation--dissipation theorem \cite{Scheel} is a Riemann--Hilbert problem \cite{AF}. 

\subsection{Kubo--Martin--Schwinger relation}

Riemann--Hilbert problems require further conditions for having unique solutions \cite{AF}. One sees this here from a simple physical argument: according to Eqs.~(\ref{gamma}) and (\ref{green0}) the dissipation $\Gamma$ does not depend on temperature, but the fluctuations $K$ should be. However, Eq.~(\ref{rh}) does not encode the temperature, so it cannot possibly be complete for determining $K$ from $\Gamma$. The relevant condition from physics is known as the Kubo--Martin--Schwinger (KMS) relation \cite{Scheel}. 

Consider the propagation of the electromagnetic field in conformal time $\tau$. Since space--time appears flat in conformal coordinates, Eq.~(\ref{conformallyflat}), the electromagnetic field evolves with free--field Hamiltonian $\hat{H}$:
%%%%%%
\begin{equation}
\hat{A}(\tau_1) = 
\hat{U}(\tau_0-\tau_1) \, \hat{A}(\tau_0)\, \hat{U}(\tau_1-\tau_0) \,,\quad
\hat{U}(\tau) = \exp\left(-\frac{\mathrm{i}}{\hbar}\, \hat{H}\right) \,.
\label{free}
\end{equation}
%%%%%%
As shown in Sec.~1.4, the quantum vacuum appears inside cosmological horizons as thermal radiation. For $\tau_1\sim\tau_0$ and $\bm{r_1}\sim\bm{r_0}$ the two points of the correlation function $f(z)$ lie within a common horizon. Consequently, one can assume as quantum state a thermal state with conformal Gibbons--Hawking temperature given by Eq.~(\ref{conftemp}) and density matrix \cite{LL5}
%%%%%%
\begin{equation}
\hat{\rho} = \frac{1}{Z}\, \mathrm{e}^{-\beta \hat{H}} \,,\quad \beta = \frac{1}{k_\mathrm{B}\Theta} = \frac{2\pi}{\hbar\dot{n}} \,.
\label{thermal}
\end{equation}
%%%%%%
Assume further that the time--derivative $\dot{n}$ of the refractive index, and hence the temperature, is taken at the time $t$ that corresponds to the conformal time $\tau$ between $\tau_1$ and $\tau_0$ as defined in Eqs.~(\ref{tau}) and (\ref{tau10}). 

Consider now the complex conjugate of $f$ on the real axis, $f^*(\sigma)$. As the field operators $\hat{A}$ are Hermitian, $f^*(\sigma)$ is proportional to $\langle \hat{A}_0 \hat{A}_1 \rangle$ and so, from Eqs.~(\ref{tau10}), (\ref{fdef}) and (\ref{thermal}) follows
%%%%%%
\begin{eqnarray}
f^*(\sigma) &\propto& \mathrm{tr}\left\{\mathrm{e}^{-\beta \hat{H}}\hat{A}_0(\tau_0)\,\hat{A}_1(\tau_1)\right\} \nonumber\\
&=&  \mathrm{tr}\left\{\mathrm{e}^{-(\beta/2) \hat{H}}\hat{A}_0(\tau_0)\,\hat{A}_1(\tau_1)\,\mathrm{e}^{-(\beta/2) \hat{H}}\right\} \nonumber\\
&=& \mathrm{tr}\left\{\hat{A}_0(\tau_0+\frac{\mathrm{i}\hbar\beta}{2})\,\mathrm{e}^{-\beta \hat{H}}\hat{A}_1(\tau_1-\frac{\mathrm{i}\hbar\beta}{2})\right\} \nonumber\\
&=& \mathrm{tr}\left\{\mathrm{e}^{-\beta \hat{H}}\hat{A}_1(\tau_1-\frac{\mathrm{i}\hbar\beta}{2})\,\hat{A}_0(\tau_0+\frac{\mathrm{i}\hbar\beta}{2})\right\} \nonumber\\
&\propto&f(\sigma_*)
\end{eqnarray}
%%%%%%
with the definition
%%%%%%
\begin{equation}
\sigma_*\equiv\sigma - \frac{2\pi \mathrm{i}}{\dot{n}} \,.
\label{sigma}
\end{equation}
%%%%%%
This gives the KMS relation
%%%%%%
\begin{equation}
f(\sigma_*) = f^*(\sigma)
\label{kms}
\end{equation}
%%%%%%
that completes the Riemann--Hilbert problem of Eq.~(\ref{rh}) and establishes the fluctuation--dissipation theorem. 

\subsection{Conformal mapping}

The standard techniques for deducing the correlation $K$ of the electromagnetic field fluctuations from the classical Green functions $\Gamma$ employ Fourier transformation \cite{Scheel} or expansion in terms of temperature Green functions \cite{LL9} (Matsubara method). The techniques and their result for constant temperature are well--known textbook material, but here I develop a geometrical method that is generalizable to situations where the temperature varies with time. This will be essential for the renormalization where the Gibbons--Hawking temperature of the self--interaction depends on the time of emission. The geometrical method also has the pedagogical advantage of arriving at the known answer with minimal algebra. 

\begin{figure}[t]
\begin{center}
\includegraphics[width=20pc]{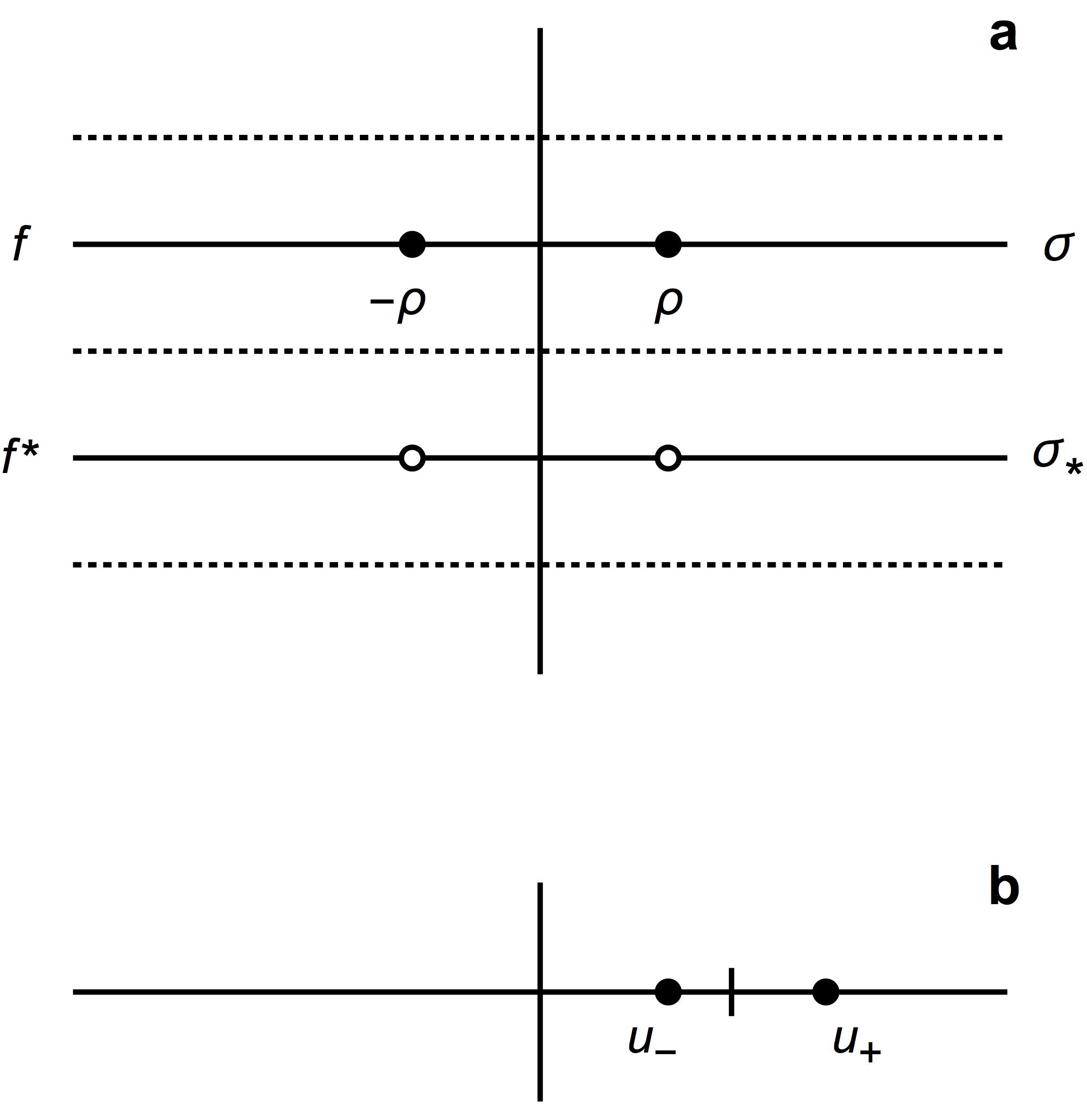}
\caption{
\small{
Analytic structure and conformal mapping. {\bf a}: diagram of the complex $z$--plane on which the field--correlation function $f(z)$ of Eq.~(\ref{fdef}) is defined. The $\sigma$--line denotes the real axis, $\sigma_*$ the line, Eq.~(\ref{sigma}), of the Kubo--Martin--Schwinger relation (\ref{kms}). On both lines, $f(z)$ is real, apart from delta--function singularities indicated by dots [with $\rho$ given by Eq.~(\ref{r})]. On the dotted lines between, below and above, $f$ is real. The horizontal lines (dotted or not) cut the plane into strips where the function alternates between $f$ on one strip and its complex conjugate $f^*$ (according to the Schwarz reflection principle \cite{AF}). The function $f$ is thus periodic in twice the strip width, apart from delta--function singularities, and can be conformally mapped to one complex $w$--plane by the exponential map, Eq.~(\ref{exp}). {\bf b}: diagram of the $w$--plane. Here $f(z(w))$ is analytic, apart from two poles (dots) at $u_\pm$ that are inverse to each other around unity (dash). From the Kramers--Kronig relation on the $w$--plane follows the field--correlation function.
 }
\label{riemann}}
\end{center}
\end{figure}

Figure~\ref{riemann}a illustrates the structure of the complex plane on which the function $f(z)$ is defined in Eq.~(\ref{fdef}) with KMS relation (\ref{kms}) for a temperature that only depends on the average conformal time $\tau$ of emission and reception, but not on the difference $\sigma$. In this case the temperature plays the role of an external parameter, which is equivalent to the standard, known case of constant temperature. On the real axis, $z=\sigma$, the imaginary part of $f$ consists solely of two delta--function singularities at $\pm\rho$ with $\rho=r/c$ [according to Eqs.~(\ref{gamma}), (\ref{green0}), (\ref{r}) and (\ref{rh})]. On the line $z=\sigma_*$ of Eq.~(\ref{sigma}) parallel to the real axis, $f$ must be the complex conjugate, as the KMS relation (\ref{kms}) requires. So here $f$ is real as well, apart from two delta--function singularities (Fig.~\ref{riemann}a). Symmetry requires that exactly between the two parallel lines, at $z=\sigma-\mathrm{i}\pi/\dot{n}$, the function $f$ is real.  According to the Schwarz reflection principle \cite{AF}, in the strip between $z=\sigma$ and $z=\sigma+\mathrm{i}\pi/\dot{n}$ the function $f$ is the complex conjugate of the strip below. Repeated applications of the Schwarz reflection principle show that the pattern of $f$ and its complex conjugate repeats itself in all the strips below and above. One sees that $f$ is periodic in $2\pi \mathrm{i}/\dot{n}$ apart from delta--function singularities. 

%%%
%%%

Let me therefore transform the variable of $f$ from $z$ to $w$ by the exponential map \cite{Needham}:
%%%%%%
\begin{equation}
w = \mathrm{e}^{\dot{n} z} \,,\quad w = u+\mathrm{i}v 
\label{exp}
\end{equation}
%%%%%%
that collects all the periods in one complex plane (Fig.~\ref{riemann}b). On the $w$--plane the Green function $G_\pm$ of Eq.~(\ref{green0}) appears as 
%%%%%%
\begin{equation}
G_\pm =
-\frac{1}{4\pi c\rho}\,\frac{\mathrm{d}u_\pm}{\mathrm{d}\rho}\,\delta(u-u_\pm) \,,\quad
u_\pm = u(\pm\rho) \label{upm} 
\end{equation}
%%%%%%
producing via Eqs.~(\ref{gamma}) and (\ref{rh}) the delta--function singularities indicated in Fig.~\ref{riemann}b. The function $f(z(w))$ is therefore analytic on the $w$--plane, except at the two singularities. From the Hilbert transformation \cite{AF} (Kramers-Kronig relation)
%%%
\begin{equation}
\mathrm{Re} f(z(u)) = \frac{1}{\pi} \dashint_{-\infty}^{+\infty} \frac{\mathrm{Im} f(z(\chi))}{\chi-u}\,\mathrm{d}\chi
\label{hilbert}
\end{equation}
%%%
and the fluctuation--dissipation relation (\ref{rh}) with $\Gamma$ given by Eq.~(\ref{gamma}) follows that $f(z(u))$ consists of two poles, apart from the delta--function singularities in the Green function. Ignoring these contact terms and analytically continuing $f$ on the complex $w$--plane gives
%%%%%%
\begin{equation}
f = -\frac{1}{8\pi c^2\rho} \sum_\pm \frac{\mathrm{d}u_\pm}{\mathrm{d}\rho} \,\frac{1}{\pi(w-u_\pm)} 
= -\frac{1}{8\pi^2 c^2\rho}\, \partial_\rho \ln \left[(w-u_+)(w-u_-)\right] \,. 
\label{fresult}
\end{equation}
%%%%%%
From this result follows the correlation function $K$ via the fluctuation--dissipation relation of Eq.~(\ref{rh}) and the exponential map of Eq.~(\ref{exp}):
%%%
\begin{equation}
K = -\frac{1}{8\pi^2 c^2\rho}\, \partial_\rho \ln \left[(\mathrm{e}^{\dot{n}\sigma}-\mathrm{e}^{\dot{n}\rho})(\mathrm{e}^{\dot{n}\sigma}-\mathrm{e}^{-\dot{n}\rho})\right]
\end{equation}
%%%
that agrees with the known result for constant temperature \cite{Scheel} with the conformal Gibbons--Hawking temperature of Eq.~(\ref{conftemp}). In the geometrical method developed here, the problem is reduced to a Hilbert transformation (Kramers--Kronig relation) and a conformal map that encodes the temperature.

The correlation function $K$ of Eq.~(\ref{fresult}) diverges for $\sigma\rightarrow 0$ and $\rho\rightarrow 0$ that correspond to the limits $\tau_1\rightarrow\tau_0$ and $\bm{r}_1\rightarrow\bm{r}_0$ according to Eqs.~(\ref{tau10}) and (\ref{r}). In these limits,
%%%%%%
\begin{equation}
K\sim -\frac{1}{8\pi^2 c^2\rho}\, \partial_\rho \ln \left[(\sigma-\rho)(\sigma+\rho)\right] =  \frac{1}{4\pi^2c^2(\sigma^2-\rho^2)} \,.
\end{equation}
%%%%%%
Taking the time--dependent limit ($\sigma\rightarrow 0$) first --- assuming spatial dispersion --- gives the electromagnetic energy density 
%%%%%%
\begin{eqnarray}
u_\mathrm{vac} &=& \left.\frac{\hbar}{n c} \left( \frac{1}{4}\, \partial_\tau^2 - \partial_\sigma^2 - \frac{1}{\rho^2} \, \partial_\rho \rho^2 \partial_\rho \right) K\, \right|_{\sigma=0}
\label{uvack}\\
&=& \frac{\hbar c}{\pi^2 n r^4} 
\label{uvacuum}
\end{eqnarray}
%%%%%%
according to Eqs.~(\ref{uvac}) and (\ref{r}). During cosmic expansion with the metric of Eq.~(\ref{metric}) the length $r$ is not invariant, but the length $nr$ is. Setting the invariant length $nr$ to the Planck length of Eq.~(\ref{planck}) produces for the cosmological energy density $\epsilon_\mathrm{vac}=u_\mathrm{vac}/n^3$ the expression
%%%
\begin{equation}
\epsilon_\mathrm{vac} = \frac{\hbar c}{\pi^2 \ell_\mathrm{p}^4}
\end{equation}
%%%
that disagrees with the energy density of the cosmological constant of Sec.~1.5 $\epsilon_\Lambda\sim (\hbar/c) H^2/ \ell_\mathrm{p}^2$ by a factor of $c^2/(\pi \ell_\mathrm{p} H)^2\sim3\times 10^{120}$ for a Hubble constant $H$  of  about 1 per $10^{10}$ years, where it not for renormalization. 

\subsection{Causality and second order}

In renormalization, the energy density of the unphysical self--interaction of each source is subtracted from Eq.~(\ref{uvacuum}). For describing the self--interaction, we found in our previous work \cite{Grin1} on the Casimir stress in planar media that the refractive--index profile should be expanded to second order around the point of emission. Here I assume the same for time--dependent media:
%%%
\begin{equation}
n(t_1,t_0)=n(t_0)+\dot{n}(t_0)\,(t_1-t_0) + \frac{\ddot{n}(t_0)}{2}\,(t_1-t_0)^2 \,.
\label{nexpansion}
\end{equation}
%%%
Replacing the actual $n(t)$ by the $n(t_1,t_0)$ the self--interaction experiences causes two variations, in the Green functions of the dissipation $\Gamma$ and in the effective temperature. The Green functions of Eqs.~(\ref{gamma}) and (\ref{green0}) depend on the difference $\sigma$ in conformal time, defined in Eqs.~(\ref{tau}) and (\ref{tau10}):
%%%%%%
\begin{equation}
\sigma = \int_{t_0}^{t_1} \frac{\mathrm{d}t}{n} \sim \frac{t_1-t_0}{n(t_0)} + \left. \frac{(t_1-t_0)^2}{2}\,\partial_t \frac{1}{n}\right|_{t_0} +  \left. \frac{(t_1-t_0)^3}{6}\,\partial_t^2 \frac{1}{n}\right|_{t_0} \,.
\end{equation}
%%%%%%
Consider first the effect of this variation on the energy density, assuming the Gibbons--Hawking temperature to be unchanged. As
%%%
\begin{equation}
\partial_{t_0}\sigma = -\frac{1}{n(t_0)} +  \left. \frac{(t_1-t_0)^3}{6}\,\partial_t^3 \frac{1}{n}\right|_{t_0} 
\end{equation}
%%%
the variation produces an extra contribution to the energy density of Eq.~(\ref{uvac}) proportional to $(t_1-t_0)^3$ that does not diverge and, moreover, vanishes if the limit $t_1 \rightarrow t_0$ is taken first. Therefore, the only significant correction to the vacuum energy density can come from the Gibbons--Hawking temperature ---- from cosmological horizons, as argued already in Sec.~1.4 on general grounds. 

\begin{figure}[t]
\begin{center}
\includegraphics[width=20pc]{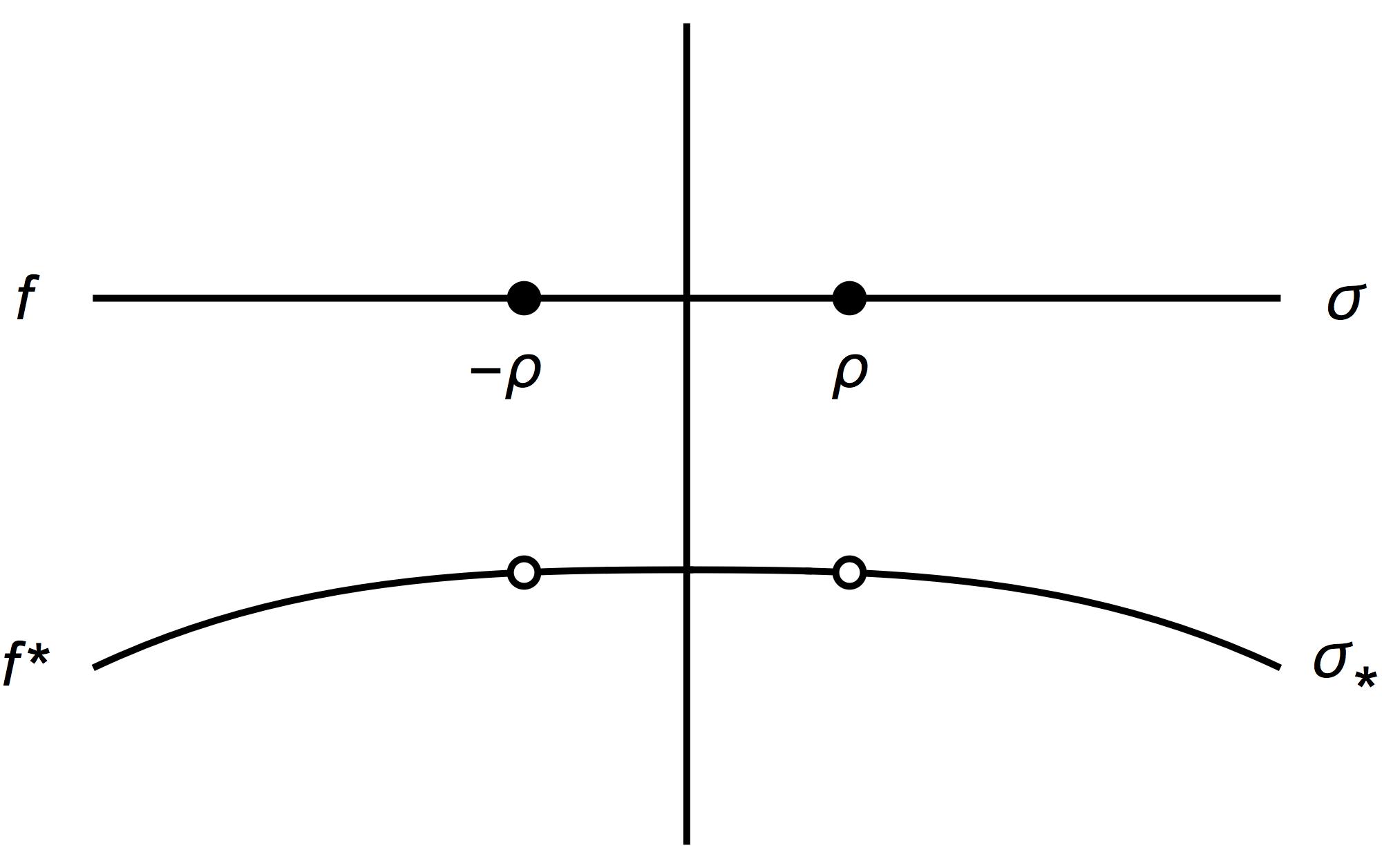}
\caption{
\small{
Kubo--Martin--Schwinger relation for the self--interaction. On the curve $z=\sigma_*$ the correlation function $f$ is the complex conjugate of the $f$ on the real axis where $z=\sigma$. The dots indicate the singularities  of $f$ (Fig.~\ref{riemann}). The $\sigma_*$ curve is given by Eqs.~(\ref{sigma}) and (\ref{expdot}). Due to causality the curve is symmetric around the imaginary axis ($\sigma_*$ depends on $|\sigma|$). The third derivative of the curve is discontinuous. This subtle discontinuity in the KMS curve is going to create the divergence in the self--interaction that may appear on cosmological scales. 
}
\label{curve}}
\end{center}
\end{figure}

There is another important point to consider: causality. Expansion (\ref{nexpansion}) tacitly assumes that $t_0$ is the time of emission and $t_1$ the time of reception, but the fluctuation--dissipation theorem draws on all times, including $t_1$ preceding $t_0$, as $\sigma$ of Eqs.~(\ref{tau}) and (\ref{tau10}) runs from $-\infty$ to $+\infty$ in Eq.~(\ref{sigma}) of the KMS relation (\ref{kms}). The refractive index effective for the self--interaction can only depend on the causal time order, {\it i.e.} on $|\sigma|$. Pictorially (Fig.~\ref{curve}) the $\sigma_*$--curve of Eq.~(\ref{sigma}) must be symmetric around the imaginary axis. Since $\dot{n}$ in Eq.~(\ref{sigma}) is already a first derivative, one needs to expand linearly from the time of emission:
%%%
\begin{equation}
\frac{1}{\dot{n}_\mathrm{eff}} = \frac{1}{\dot{n}(\tau-|\sigma|/2)} + \frac{|\sigma|}{2} \,\partial_\tau \frac{1}{\dot{n}} \,.
\label{expdot}
\end{equation}
%%%
This expansion inserted instead of $1/\dot{n}$ in Eq.~(\ref{sigma}) defines the KMS relation (\ref{kms}) for the self--interaction. Expanding Eq.~(\ref{expdot}) for small $\sigma$, 
%%%
\begin{equation}
\frac{1}{\dot{n}_\mathrm{eff}} \sim \frac{1}{\dot{n}} + \frac{\sigma^2}{8}\,\partial_\tau^2 \frac{1}{\dot{n}} - \frac{|\sigma|^3}{48}\,\partial_\tau^3 \frac{1}{\dot{n}} \,,
\label{expneff}
\end{equation}
%%%
reveals that for the KMS curve $z=\sigma_*$ of Eq.~(\ref{sigma}) the third derivative is discontinuous at $\sigma=0$ (on the imaginary axis). This third--order discontinuity may cause an additional divergence in the renormalized vacuum energy. For capturing its effect with the tools of complex analysis some help from geometry in the complex plane is needed, as follows. 

%%%
%%%

\subsection{Characterization of complex curves}

The problem of calculating the correlation function of the self--interaction is solved if one finds a conformal map that reduces it (Fig.~\ref{curve}) to the case of constant temperature (Fig.~\ref{riemann}) --- at least locally\footnote{For the correlation of the self--interaction, $f$ is not analytic as a function of $z$, strictly speaking, as the density matrix depends on $|\sigma|=|\mathrm{Re}\,z|$ via Eqs.~(\ref{thermal}) and (\ref{expdot}). However, $f$ can be made locally analytic around $z\sim\sigma_*\sim-2\pi \mathrm{i}/\dot{n}$ where it is needed in the KMS relation (\ref{kms}). This is done by representing $f$ as a function of $w$ with the help of Eq.~(\ref{zschwarz}).}. The expansion (\ref{expneff}) in Eq.~(\ref{sigma}) is manifestly non--analytic though, as the third derivative is discontinuous. The first task therefore is to parametrize a curve in the complex $z$--plane with an analytic function $z(w)$, given the value and the derivatives up to third order with respect to another, non--analytic parameterization. Then the discontinuity must be implemented with a suitable analytic function that has a branch point there. In the following I solve the first problem by identifying the parameter--invariant geometrical quantities of a complex curve up to third order, and expressing them in terms of $z(w)$.

%%%
\begin{figure}[t]
\begin{center}
\includegraphics[width=19pc]{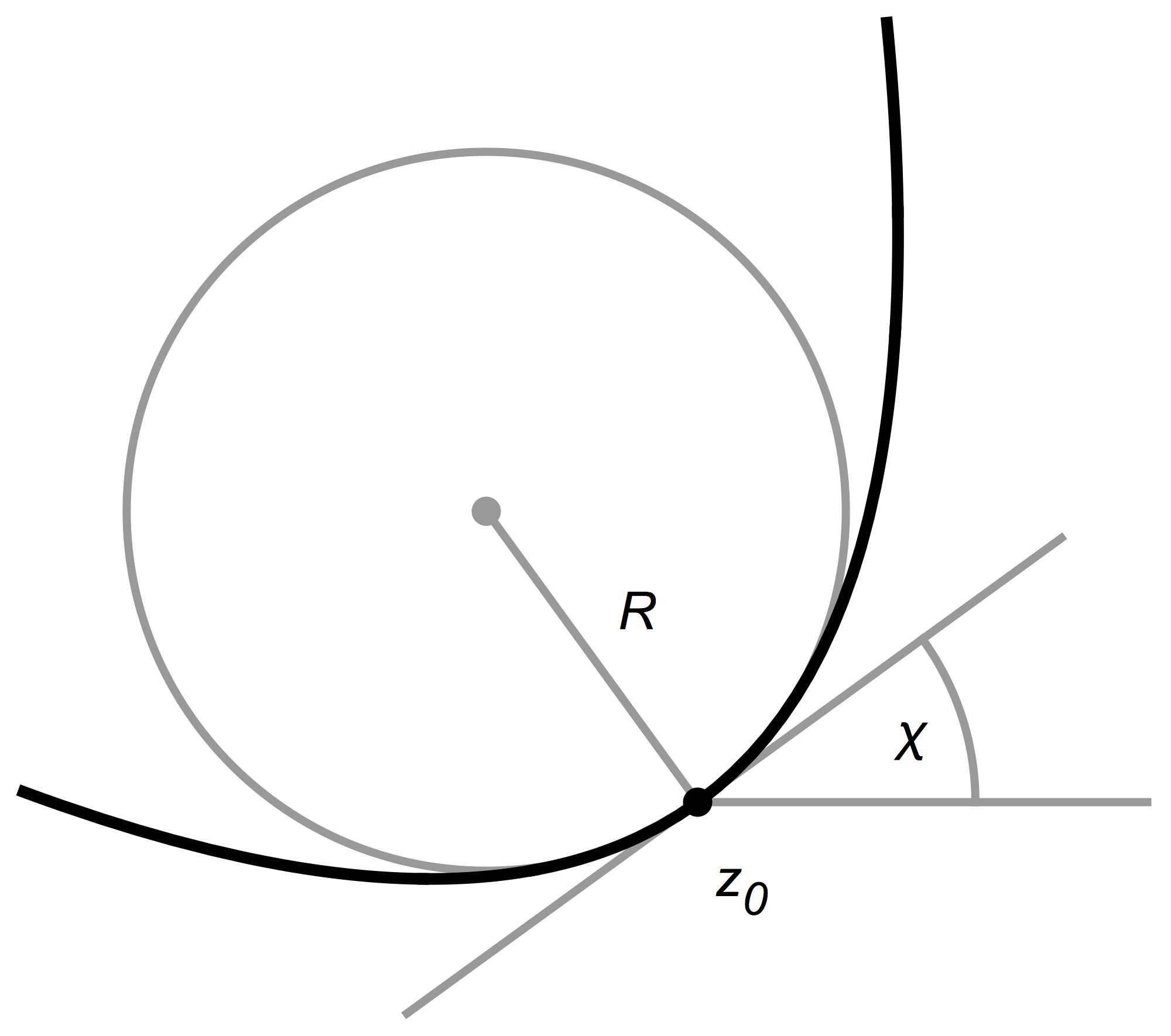}
\caption{
\small{
Characterization of curves. The black curve though the point $z_0$ is characterized to second order by the angle $\chi$ and the radius of curvature $R$. The derivative of $1/R$ with respect to the curve length characterizes the curve to third order, which is helpful for analysing the third--order discontinuity of the KMS curve in Fig.~\ref{curve}. 
}
\label{character}}
\end{center}
\end{figure}
%%%

Consider a curve in the complex plane going through the point $z_0$ (Fig.~\ref{character}). Let the curve cross $z_0$ under the angle $\chi$ with respect to the real axis:
%%%
\begin{equation}
\chi = \mathrm{arg}\, z' = \mathrm{Im} \ln z' \,.
\label{angle}
\end{equation}
%%%
Here and in the rest of this section dashes denote derivatives with respect to $w$. The next geometrical quantity is the radius of curvature $R$. One determines $R$ by differentiating the angle with respect to the curve length with increment
%%%
\begin{equation}
\mathrm{d}\ell = |\mathrm{d}z| = |z'|\, \mathrm{d}w
\end{equation}
%%%
assuming $\mathrm{d}w$ to be real ($w$ to be parallel to the real axis). As $\mathrm{d}\ell=R\, \mathrm{d}\chi$ one gets the well--known expressions
%%%
\begin{equation}
\frac{1}{R} = \frac{\mathrm{d}\chi}{\mathrm{d}\ell} = \frac{1}{|z'|}\,\mathrm{Im}\, \partial_\omega \ln z' =  \frac{1}{|z'|}\,\mathrm{Im}\, \frac{z''}{z'} \,.
\label{curvR}
\end{equation}
%%%
The point, angle and radius of curvature characterize the curve up to second order, but here I also need the third--order quantity. Let me differentiate $\mathrm{d}\chi/ \mathrm{d}\ell$ with respect to the length one more time:
%%%
\begin{eqnarray}
\frac{\mathrm{d}^2\chi}{\mathrm{d}\ell^2} &=& \frac{1}{|z'|^2}\,\mathrm{Im} \left(\partial_w \frac{z''}{z'}\right) + \frac{1}{|z'|}\left(\mathrm{Im}\,\frac{z''}{z'}\right) \partial_w \frac{1}{|z'|} \,, \nonumber\\
\partial_w \frac{1}{|z'|} &=& - \mathrm{Re}\,\frac{z'^* z''}{|z'|^2} = - \frac{1}{|z'|} \mathrm{Re}\, \frac{z''}{z'} \,.
\end{eqnarray}
%%%
As for a complex number $\mathrm{Im}\,Z^2=2(\mathrm{Im}\, Z)(\mathrm{Re}\, Z)$ I combine the terms of $\mathrm{d}^2\chi/ \mathrm{d}\ell^2$ in the compact expression
%%%
\begin{equation}
\frac{\mathrm{d}^2\chi}{\mathrm{d}\ell^2}  =  \frac{1}{|z'|^2}\,\mathrm{Im}\,\{z,w\}
\label{chi2}
\end{equation}
%%%
in terms of the Schwarzian derivative \cite{AF} 
%%%
\begin{equation}
\{z,w\} \equiv \left(\frac{z''}{z'}\right)' - \frac{1}{2}\left(\frac{z''}{z'}\right)^2 \,.
\label{schwarz}
\end{equation}
%%%
The imaginary part of the Schwarzian thus establishes the required third--order quantity. Since both $\chi$ and $\ell$ are parameter--invariant, $\mathrm{d}^2\chi/ \mathrm{d}\ell^2$ is invariant, too.

\subsection{Schwarzian discontinuity}

In the case considered here, the KMS curve (\ref{sigma}) of the self--interaction with Eq.~(\ref{expneff}) as effective refractive index, the curve is characterized for $\sigma\sim 0$ by Eqs.~(\ref{angle}) and (\ref{curvR}) with parameter $w=\sigma$:
%%%
\begin{equation}
z_0= - \frac{2\pi \mathrm{i}}{\dot{n}} \,,\quad \chi_0 = 0 \,,\quad \left.\frac{\mathrm{d}\chi}{\mathrm{d}\ell}\right|_0 = -\frac{\pi}{2}\,\partial_\tau^2 \frac{1}{\dot{n}}
\end{equation}
%%%
and the third--order quantity from Eqs.~(\ref{chi2}) and (\ref{schwarz}):
%%%
\begin{equation}
\left.\frac{\mathrm{d}^2\chi}{\mathrm{d}\ell^2}\right|_{\pm 0} = \pm \frac{\pi}{4}\,\partial_\tau^3 \frac{1}{\dot{n}}
\label{chi2expr}
\end{equation}
%%%
that changes sign at $\sigma = 0$. As the next step I need to express $\mathrm{d}^2\chi/ \mathrm{d}\ell^2$ in Eqs.~(\ref{chi2}) and (\ref{schwarz}) by the imaginary part of an analytic function --- an imaginary part that changes sign. This function is the logarithm. For keeping the notation uncluttered it is wise to move the parameter $w$ to the real axis. I thus require
%%%
\begin{equation}
\{z,w\} = \gamma + \delta \ln w 
\label{schwarzeq}
\end{equation}
%%%
with complex constant $\gamma$ and real constant $\delta$. I choose the scale of the parameter $w$ such that $z'(0)=1$ and get from Eq.~(\ref{chi2}) of $\mathrm{d}^2\chi/\mathrm{d}\ell^2$ and the $\pi$ jump of the imaginary part of the logarithm:
%%%
\begin{equation}
\mathrm{Im}\,\gamma = \left.\frac{\mathrm{d}^2\chi}{\mathrm{d}\ell^2}\right|_{+0} \,,\quad \pi\delta = -2\, \left.\frac{\mathrm{d}^2\chi}{\mathrm{d}\ell^2}\right|_{+0} \,.
\label{deltaconstant}
\end{equation}
%%%
Assuming for $w\sim 0$ (for $\sigma\sim 0$) that $z\sim z_0+w+\mathrm{i}\alpha w^2$ with real constant $\alpha$ gives
%%%
\begin{equation}
\mathrm{Re}\,\gamma = 2\alpha^2 \,,\quad \alpha = \left.\frac{\mathrm{d}\chi}{\mathrm{d}\ell}\right|_{+0} \,.
\end{equation}
%%%
The Schwarzian discontinuity appears in higher--order and logarithmic terms of $z(w)$. Consider
%%%
\begin{equation}
z \sim z_0 + w + \mathrm{i}\alpha w^2 + \frac{\delta}{6}\, w^3 \ln w + \left(\frac{\gamma}{6} - \frac{11}{36}\,\delta\right) w^3 \,.
\label{zschwarz}
\end{equation}
%%%
This expression is consistent with the principal behavior of $z\sim z_0+w+\mathrm{i}\alpha w^2$ for $w\sim 0$. It satisfies
%%%
\begin{equation}
\left. z\right|_0 = z_0 \,,\quad \left. z'\right|_0 = 1 \,,\quad \left. z''\right|_0 = 2\mathrm{i}\alpha \,,
\end{equation}
%%%
and in the third order
%%%
\begin{equation}
z''' = \gamma + \delta \ln w \,.
\end{equation}
%%%
Expression (\ref{zschwarz}) thus fulfills Eq.~(\ref{schwarzeq}) for the Schwarzian of Eq.~(\ref{schwarz}) and $w\sim 0$. The problem is solved. 

\subsection{Self--energy density}

Formula (\ref{zschwarz}) describes the $\sigma_*$ curve of the KMS relation (Fig.~\ref{curve}) as an analytic function of $w$ evaluated at the line $w=\mathrm{real}$, {\it i.e.} at the real axis in the $w$--plane (Fig.~\ref{riemann}b). Here the dominant, singular contributions to the correlation $f$ are captured by the open Cauchy integral of Eq.~(\ref{hilbert}) because the closing integration contour required by Cauchy's theorem \cite{AF} will not produce a term that diverges on the real $w$--axis. One thus obtains Eq.~(\ref{fresult}) for $f$ as well, yet this time not as an exact solution, but as describing the dominant behavior near the $w$--line that corresponds to the $\sigma_*$ curve in the $z$--plane. On this curve, $f$ is real (apart from delta--function singularities one can ignore in the point--splitting method). In the KMS relation (\ref{kms}) this real $f$ is projected onto the real $z$--axis (Fig.~\ref{curve}) where
%%%%%%
\begin{equation}
\sigma=\mathrm{Re}\,z \sim u +\frac{\delta}{6}\, u^3 \ln |u| \quad\mbox{for}\quad u\sim 0 \,.
\end{equation}
%%%%%%
One solves for $u$, 
%%%%%%
\begin{equation}
u \sim \sigma -\frac{\delta}{6}\, \sigma^3 \ln |\sigma| \quad\mbox{for}\quad \sigma\sim 0 \,,
\end{equation}
%%%%%%
and applies the so--transformed $u$ in the dominant contribution to the correlation function:
%%%%%%
\begin{equation}
K= -\frac{1}{8\pi^2 c^2\rho}\, \partial_\rho \ln \left[(u-u_+)(u-u_-)\right] 
\label{ku}
\end{equation}
%%%%%%
according to Eqs.~(\ref{rh}) and (\ref{fresult}) with $u_\pm$ defined in Eq.~(\ref{upm}). Expression (\ref{ku}) is then inserted in Eq.~(\ref{uvack}) for calculating the energy density $u_\mathrm{vac}^0$ of the self--interaction. Taking the limit $\sigma\rightarrow 0$ first ($t_1\rightarrow t_0$) and expressing $\rho$ as $r/c$ gives 
%%%%%%
\begin{equation}
u_\mathrm{vac}^0 = \frac{\hbar c}{\pi^2 n r^4} - \frac{\hbar\delta}{6\pi^2 cn r^2} + \mathrm{O}[(\ln\rho)^2] \,.
\label{uvacuum0}
\end{equation}
%%%%%%
Consequently, in the limit of purely spatial dispersion, the self--energy density of the electromagnetic vacuum compensates for the vacuum energy density of Eq.~(\ref{uvacuum}) apart from a term diverging with an inverse length squared. This is required (Sec.~1.5) for a realistic cosmological constant. Assuming temporal dispersion and hence attempting to take the limit $\rho\rightarrow 0$ first ($\bm{r}_1\rightarrow\bm{r}_0$) produces an additional $\sigma^{-2}\ln\rho$ singularity that prevents the convergence of the limit $\rho\rightarrow 0$. In the case of spatial dispersion, I obtain from Eqs.~(\ref{epsvac}),  (\ref{uvacuum}), (\ref{chi2expr}), (\ref{deltaconstant}) and (\ref{uvacuum0}) for the cosmological energy density:
%%%%%%
\begin{equation}
\epsilon_\mathrm{vac} = - \frac{\hbar}{12\pi^2 cr\, n^4}\, \partial_\tau^3 \frac{1}{\dot{n}} \,.
\label{epsvacflat}
\end{equation}
%%%%%%
Applying the definitions (\ref{tau}) and (\ref{hubble}) for the conformal time $\tau$ and the Hubble constant $H$, and differentiating thrice gives
%%%%%%
\begin{equation}
\partial_\tau^3\frac{1}{\dot{n}} = (n\partial_t)^3 \frac{1}{\dot{n}} = n^2\left(\partial_t^3\frac{1}{H} + H\partial_t^2\frac{1}{H}\right) \,.
\end{equation}
%%%%%%
Setting the invariant length $n r$ to the Planck length $\ell_\mathrm{p}$ defined in Eq.~(\ref{planck}) I thus arrive at the main result of this paper, Eqs.~(\ref{delta}) and (\ref{result}), with the constant 
%%%%%%
\begin{equation}
\alpha_\Lambda=\frac{1}{9\pi} \approx 10^{-3}\,.
\end{equation}
%%%%%%
The coefficient $\alpha_\Lambda$ is not precisely defined, as the exact behavior of the electromagnetic field near the Planck scale is yet unknown. In particular, there is no precise cut--off known. Dividing the cut--off length by a number $s$ multiplies $\alpha_\Lambda$ by $s^2$, which leads to uncertainty in the value of $\alpha_\Lambda$. However, the approximate order of magnitude of $\alpha_\Lambda$ is given, and this figure makes the electromagnetic vacuum energy relevant on cosmological scales. For a spatially flat, empty universe, the electromagnetic vacuum contributes to the cosmological constant $\Lambda$ according to Eq.~(\ref{lambda}). Here $\Lambda$ is predominantly given by the Hubble constant with a quantum correction to de--Sitter expansion. It appears natural to assume that the other fields of the Standard Model act in the same way, albeit with different coefficients. Given the present state of the theory, the total $\alpha_\Lambda$ is best inferred from empirical data in astronomy. 

\section{Curvature}

So far I considered a spatially flat cosmology, because this is a good approximation for the universe at present \cite{CurveMeas} and because this is the mathematically simplest case for developing the techniques to calculate the cosmological constant in Lifshitz theory. However, space is probably slightly curved with the more general space--time metric
%%%%%%
\begin{equation}
\mathrm{d}s^2=c^2\mathrm{d}t^2-n^2\mathrm{d}\ell_0^2
\label{metricgeneral}
\end{equation}
%%%%%%
instead of Eq.~(\ref{metric}) and $\mathrm{d}\ell_0^2$ depending on the curvature of three--dimensional space in co--moving coordinates. Space is expanding in time with factor $n(t)$. The spatial curvature of the universe had been most prominent closely after the beginning, before cosmic expansion had reduced most of its effect. This section concludes the calculations by extending them to the two classic cosmological models with spatial curvature \cite{LL2,Friedman1,Friedman2}: homogeneous and isotropic space with either positive or negative curvature. 

\subsection{Homogeneous and isotropic space}

\begin{figure}[t]
\begin{center}
\includegraphics[width=16pc]{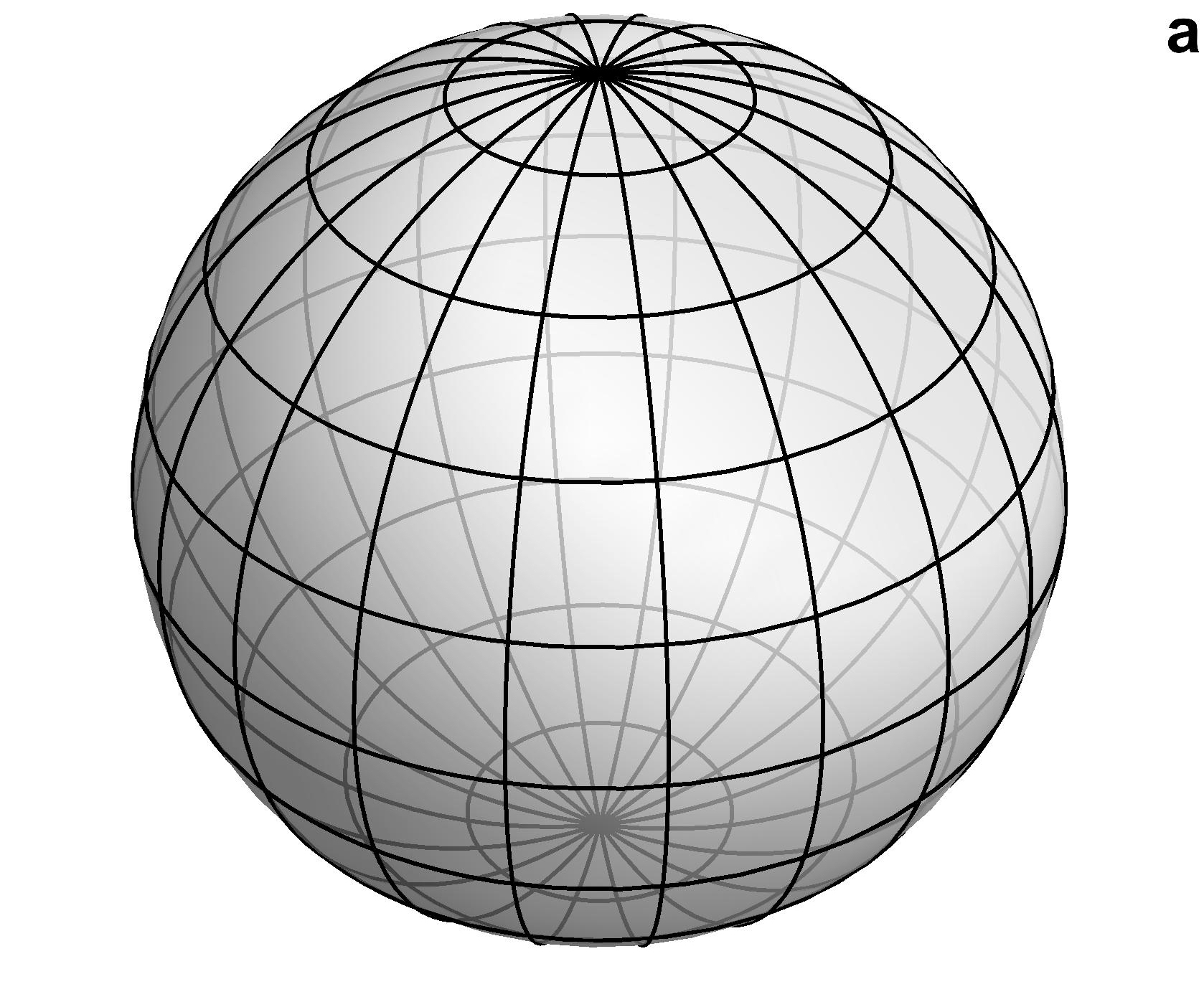}
\hspace{2pc}
\includegraphics[width=16pc]{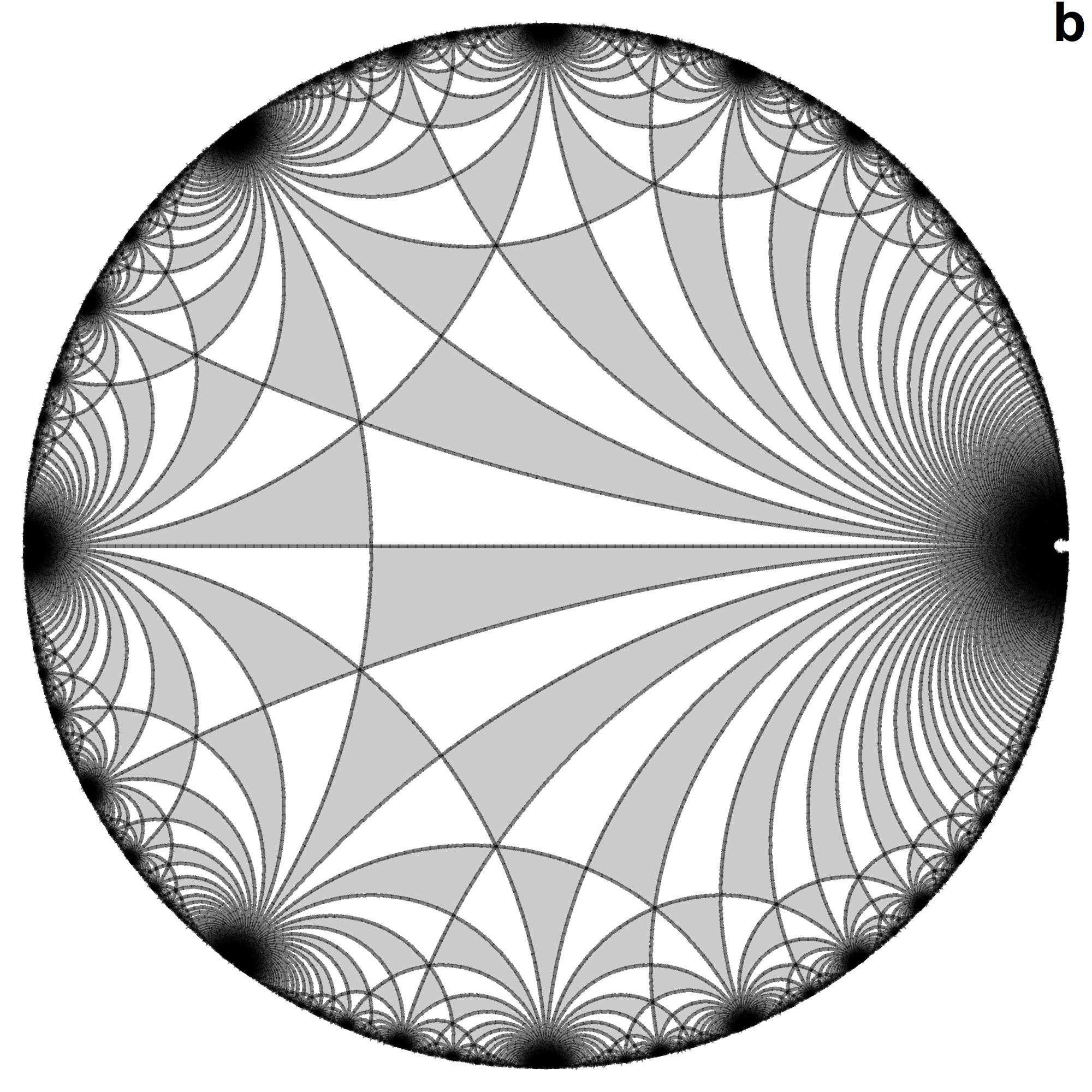}
\caption{
\small{
Curved spaces in cosmology. On cosmological scales, space is homogeneous and isotropic. {\bf a}: The positively--curved surface of a sphere or hypersphere is such a space, as all points and all directions are equal. Light propagates along great circles (lines of longitude) with orthogonal phase fronts (lines of latitude). The sphere or hypersphere is projected to co--moving coordinates by stereographic projection (Fig.~\ref{rotation}). {\bf b}: In stereographic projection, a sphere with purely imaginary radius appears as a hyperbolic space with negative curvature, the Poincar\'{e} disk. Here light propagates along circles orthogonal to the rim of the disk. The figure shows the tiling of the Poincar\'{e} disk by circles of light in the Klein invariant \cite{Erdelyi} with infinite structures near the rim: infinite space contained in a finite disk. 
}
\label{shapes}}
\end{center}
\end{figure}

Consider the spatial part of the expanding universe in co--moving coordinates and require, in accordance with empirical fact \cite{Survey}, that on cosmological scales space is homogeneous and isotropic. Such a space is maximally symmetric \cite{Zee} and there are only three possibilities: positive, negative and zero constant curvature. The space of constant positive curvature is equivalent \cite{LL2} to the three--dimensional surface of the four--dimensional hypersphere (Fig.~\ref{shapes}a). For describing the hypersurface one can use hyperspherical coordinates \cite{LL2,LeoPhil} in analogy to spherical coordinates for the two--dimensional surface of the three--dimensional sphere. Alternatively, one can project the surface to the plane or the hypersurface to three--dimensional space by stereographic projection \cite{LeoPhil,Needham}. As the stereographic projection is conformally invariant \cite{LeoPhil,Needham}, the spatial metric is conformally flat:
%%%%%%
\begin{equation}
\mathrm{d}\ell_0^2 = \nu^2 \,(\mathrm{d}x^2+\mathrm{d}y^2+\mathrm{d}z^2)
\label{spacemetric}
\end{equation}
%%%%%%
in Cartesian coordinates $\{x,y,z\}$. Isotropy requires that $\nu$ can only be a function of the radius $r$ with $r^2=x^2+y^2+z^2$. The prefactor $\nu$ acts like a spatial refractive--index profile that, together with the time--dependent expansion factor $n$, gives the total refractive index $n \nu$. Assuming that hyperspace is flat and Euclidean leads \cite{LeoPhil,Luneburg} to a refractive--index profile well--known in optics \cite{BornWolf}: Maxwell's fish eye \cite{LeoPhil,Maxwell}. The profile depends on a length scale $a$ that describes the radius of curvature, or, equivalently, the radius of the hypersphere. Replacing $a$ by a purely imaginary $\mathrm{i}a$ transmutes the positively--curved hypersurface into a negatively--curved space \cite{LL2}: Friedman's model of a cosmology with negative spatial curvature \cite{Friedman2}.

%%%
%%%

Maxwell's fish eye \cite{Maxwell} with either real or imaginary radial scale thus describes the two cosmological models with constant curvature \cite{Friedman1,Friedman2}. Let me write the known $\nu$ \cite{LeoPhil} as
%%%%%%
\begin{equation}
\nu = \frac{2}{1+\mathrm{k}\, r^2/a^2}
\label{fish}
\end{equation}
%%%%%%
where $\mathrm{k}\in\{-1,+1\}$ indicates negative ($-1$) or positive ($+1$) curvature, respectively. One obtains for the three--dimensional curvature scalar \cite{LeoPhil}:
%%%%%%
\begin{equation}
P = -\frac{4(\nabla^2 \nu)}{\nu^3} + \frac{2(\nabla \nu)^2}{\nu^4} = \frac{6\mathrm{k}}{a^2} \,.
\label{curvaturescalar}
\end{equation}
%%%%%%
Replacing $a$ by $\mathrm{i}a$ switches indeed between the two cases in the refractive--index profile of Eq.~(\ref{fish}) and flips the sign of the curvature scalar in Eq.~(\ref{curvaturescalar}). In the case of negative curvature, the profile ends at radius $r=a$ where $\nu$ diverges, but as the length increment $\mathrm{d}\ell_0$ grows without limit for $r\rightarrow a$ this hyperbolic space encloses infinitely large distances and has no end: it is infinite and open. Infinite space is contained in a finite coordinate sphere (Fig.~\ref{shapes}b). The opposite holds in the case of positive curvature: the infinitely extended coordinates describe only a finite space, the three--dimensional surface of the hypersphere, in analogy to the surface of the sphere (Fig.~\ref{shapes}a). 

\begin{figure}[t]
\begin{center}
\includegraphics[width=24pc]{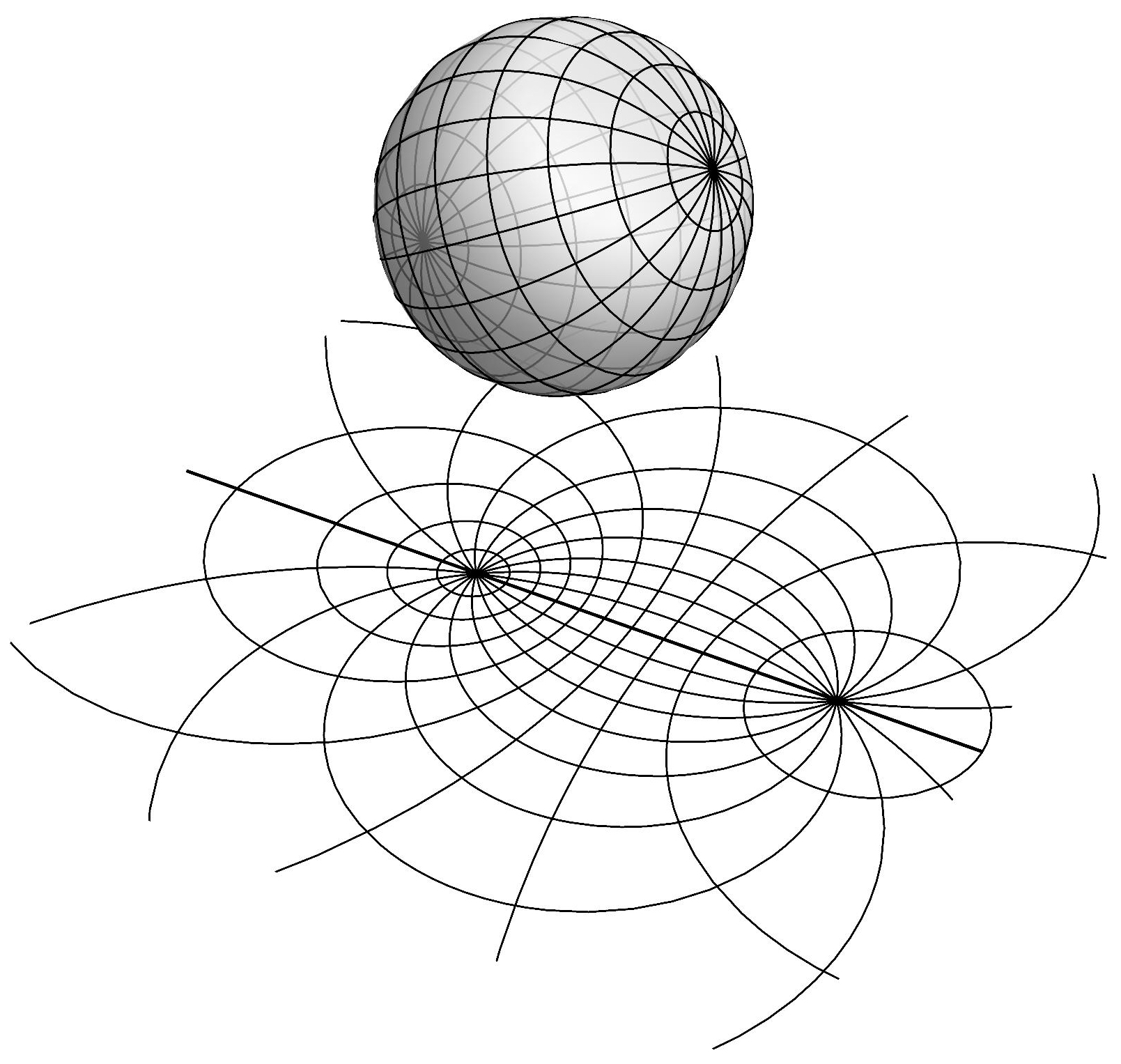}
\caption{
\small{
Rotation and M\"{o}bius transformation. Rotating the sphere appears in stereographic projection as the M\"{o}bius transformation shown below \cite{LeoPhil,Needham}. The transformed grid shows the light rays and phase fronts in the projected plane --- in Maxwell's fish eye \cite{Maxwell} that describes a universe with positive spatial curvature. The stereographic projection from a purely imaginary sphere gives the circles of light in hyperbolic space of negative curvature (Fig.~\ref{shapes}b). 
}
\label{rotation}}
\end{center}
\end{figure}

When viewed as a sphere or hypersphere, it also becomes apparent why the inhomogeneous refractive--index profile of Maxwell's fish eye describes a homogeneous space. All points on the surface of the sphere or the hypersphere are equal; these surfaces are homogeneous spaces. However, in the stereographic projection \cite{Needham,LeoPhil} a Pole is chosen from which the projection is taken, say the North Pole. The projection creates a rotationally symmetric but radially inhomogeneous profile of the line element that appears as the refractive--index profile of Eq.~(\ref{fish}). Yet any other point may become Pole simply by rotating it to Pole position (Fig.~\ref{rotation}). In stereographic projection, the rotation appears as a M\"{o}bius transformation \cite{LeoPhil,Needham} (Fig.~\ref{rotation}). Equipped with these properties, the metric (\ref{spacemetric}) with the profile of Eq.~(\ref{fish}) sets the spatial scene in the expanding universe. 

%%%
%%%

\subsection{Scalar Green function}

The calculation of the renormalized vacuum energy proceeds as in Sec.~2, except that the reduction of the electromagnetic field to two scalar polarizations is not as straightforward as in Sec.~2.3. One should regard the electromagnetic vector potential $\hat{\bm{A}}$ as a vector and consider, instead of scalar correlations and Green functions, correlation and Green bi--tensors \cite{LeoPhil}. Appendix B shows however that in the energy density only scalar Green functions $G_\pm$ appear. Each  $G_\pm$ represents the two polarizations of the electromagnetic field contributing equally to the energy. The retarded (+) and advanced (-) Green function $G_\pm$ obeys the wave equation
%%%%%%
\begin{equation}
\left(D_i D^i  - \frac{1}{c^2}\partial_\sigma^2 -\frac{p}{6}\right) G_\pm = \frac{1}{\nu^3}\,\delta(\sigma)\,\delta(\bm{r}_1-\bm{r}_0)
\label{greencurveq}
\end{equation}
%%%%%%
with $\sigma=\tau_1-\tau_0$, the curvature scalar $P$ given by Eq.~(\ref{curvaturescalar}), and $D_i D^I$ denoting the Laplacian \cite{LL2,LeoPhil} in the spatial geometry defined by Eq.~(\ref{spacemetric}):
%%%%%%
\begin{equation}
D_i D^i = \frac{1}{\nu^3}\nabla\cdot \nu \nabla \,.
\end{equation}
%%%%%%
The Green functions $G_\pm$ describe conformally--coupled scalar waves \cite{BD}, a reminiscence of the conformal invariance of Maxwellian electromagnetism \cite{LeoPhil}.

The solution of Eq.~(\ref{greencurveq}) was found in the context of perfect optical imaging \cite{LPfish}:
%%%%%%
\begin{equation}
G_\pm = - \frac{1}{4\pi r' \nu(r')}\,\delta\left(\pm\frac{1}{c}\int_0^{r'}\nu(r)\,\mathrm{d}r - \sigma\right)
\label{greencurv}
\end{equation}
%%%%%'
in terms of the M\"{o}bius--transformed propagation radius
%%%%%%
\begin{equation}
r'= \frac{|\bm{r}_1-\bm{r}_0|}{\sqrt{1+2\mathrm{k}\,\bm{r}_1\cdot\bm{r}_0\,a^{-2} +r_1^2\, r_0^2\,a^{-4}}} \,.
\label{rprime}
\end{equation}
%%%%%'
Equations (\ref{greencurv}) and (\ref{rprime}) describe a flash of light emitted or received at the point $\bm{r}_0$ on the hypersphere (Fig.~\ref{rotation}) or its hyperbolic incarnation with imaginary radial scale $a$. The flash fans out radially on the hypersurface, which appears in stereographic projection as the M\"{o}bius--transformed $r'$ of Eq.~(\ref{rprime}). The prefactor in Eq.~(\ref{greencurv}) accounts for the conservation of energy and the de--focusing and focusing along propagation. Note that Eqs.~(\ref{greencurv}) and (\ref{rprime}) describe an exact solution of geometrical optics \cite{LeoPhil,BornWolf}: there is no scattering in homogeneous and isotropic space. 

\subsection{Correlation function}

Having reduced the problem of electromagnetism in curved homogeneous and isotropic space to the scalar Green functions of Eqs.~(\ref{greencurv}) and (\ref{rprime}) I apply the mathematical machinery of Sec.~2 with a few minor modifications. Let me here define $\rho$ as
%%%%%%
\begin{equation}
\rho=\nu(r')\,\frac{r'}{c}
\label{rho}
\end{equation}
%%%%%'
such that the prefactor of $G_\pm$ in Eq.~(\ref{greencurv}) takes the same form as the prefactor in Eq.~(\ref{green0}). The delta--function singularities of Eq.~(\ref{greencurv}) lie at
%%%%%%
\begin{equation}
\sigma_\pm = \pm \frac{1}{c} \int_0^{r'} \nu(r)\,\mathrm{d}r 
\end{equation}
%%%%%'
instead of $\pm\rho$ in flat space. One obtains from Eq.~(\ref{fish}):
%%%%%%
\begin{equation}
\sigma_+ = \frac{2a}{c}
\begin{cases}
\mathrm{artanh}(r'/a) = \mathrm{arsinh}(c\rho/a) :& \mathrm{k}=-1\\
\mathrm{arctan}(r'/a) = \mathrm{arcsin}(c\rho/a) :& \mathrm{k}=+1
\end{cases}
\end{equation}
%%%%%'
for the two cases of curvature. With these modifications, the Green function of Eq.~(\ref{greencurv}) has exactly the same form as the flat--space Green function of Eq.~(\ref{green0}). According to the fluctuation--dissipation theorem, the correlation function $K$ is given by the Green functions and the temperature, so $K$ is also the same as in Eq.~(\ref{ku}) but here with $u_\pm=u(\sigma_\pm)$. The next step is the calculation of the energy density. 

\subsection{Energy density}

For calculating the energy density of the quantum vacuum I take advantage of the fact that in homogeneous and isotropic space the electric and the magnetic energy densities are equal \cite{LeoSimpson}. The electric energy density is calculated in Appendix B; twice of it gives the vacuum energy density:
%%%%%%
\begin{equation}
u_\mathrm{vac} = \left. 2\, \frac{\hbar c \nu}{n}\,\nabla_1\cdot\nabla_0 K \right|_{\sigma=0} \,.
\label{uvacc}
\end{equation}
%%%%%'
One obtains from Eq.~(\ref{rho}) for $\bm{r}_1\sim\bm{r}_0$:
%%%%%%
\begin{equation}
\nabla_1\cdot\nabla_0 K(\rho) = (\nabla_1\rho)\cdot(\nabla_0\rho)\, \partial_\rho^2 K + (\nabla_1\cdot\nabla_0\rho)\, \partial_\rho K 
\sim - \nu^2\left(\partial_\rho^2  K + \frac{2}{\rho}\,\partial_\rho K\right) ,
\end{equation}
%%%%%'
which produces in the vacuum energy density of Eq.~(\ref{uvacc}) twice the spatial term as in Eq.~(\ref{uvack}), apart from $\nu^3$ in the pre\-factor. The temporal derivatives $\partial_\sigma^2 K$ give the same as the spatial term and the $\tau$--derivatives do not produce a diverging contribution. Consequently, one obtains the same dominant energy densities as in Eqs.~(\ref{uvacuum}) and (\ref{uvacuum0}), apart from the factor $\nu^3$ and with $r$ replaced by $c\rho$ and $\rho$ given by Eq.~(\ref{rho}). The asymptotics
%%%%%%
\begin{equation}
\rho\sim \nu(r)\,\frac{|\bm{r}_1-\bm{r}_0|}{c}
\end{equation}
%%%%%'
for $\bm{r}_1\sim\bm{r}_0$ relates $\rho$ to coordinate differences $|\bm{r}_1-\bm{r}_0|$ in the point--splitting limit. Setting the invariant length $n\nu\,|\bm{r}_1-\bm{r}_0|$ to the Planck length of Eq.~(\ref{planck}) gives exactly the same cosmological energy density $\epsilon_\mathrm{vac}$ as in Eq.~(\ref{epsvacflat}). Consequently, the main result of this paper, Eqs.~(\ref{delta}) and (\ref{result}), holds also in the case of spatial curvature. 

\section{Conclusion}

Lifshitz theory in homogeneous and isotropic space with time--dependent refractive index predicts the cosmological energy density of the quantum vacuum in Eqs.~(\ref{delta}) and (\ref{result}). The trace anomaly \cite{Wald} of the vacuum energy gives the cosmological constant $\Lambda$. Here $\Lambda$ appears as a contribution to the total energy density $\epsilon$ and pressure $p$ in addition to the $\epsilon_\mathrm{m}$ and $p_\mathrm{m}$ of matter and radiation, and the $\epsilon_\mathrm{vac}$ and $p_\mathrm{vac}=\epsilon_\mathrm{vac}/3$ of the quantum vacuum itself:
%%%%%%
\begin{equation}
\epsilon=\epsilon_\mathrm{m} + \epsilon_\mathrm{vac} + \epsilon_\Lambda \,,\quad p=p_\mathrm{m} + \frac{1}{3}\,\epsilon_\mathrm{vac} - \epsilon_\Lambda \,.
\label{state}
\end{equation}
%%%%%'
From the Friedman equation \cite{LL2} in spaces of negative curvature ($\mathrm{k}=-1$), zero curvature ($\mathrm{k}=0$) and positive curvature ($\mathrm{k}=+1$) with radius $a$,
%%%%%%
\begin{equation}
H^2 + \frac{\mathrm{k}\,c^2}{n^2a^2} = \frac{8\pi G}{3c^2}\,\epsilon \,,
\label{f1full}
\end{equation}
%%%%%'
and the conservation of energy and momentum expressed in Eq.~(\ref{f2}) follows the equation of motion for the universe on cosmological scales:
%%%%%%
\begin{equation}
\dot{H} - \frac{\mathrm{k}\,c^2}{n^2a^2} = 4\alpha_\Lambda\Delta - \frac{8\pi G}{c^2}\,(\epsilon_\mathrm{m}+p_\mathrm{m})
\label{cosmicmotion}
\end{equation}
%%%%%'
where $H$ denotes the Hubble constant defined in Eq.~(\ref{hubble}), $G$ the gravitational constant and $c$ the speed of light in vacuum. The term $\Delta$ is given by Eq.~(\ref{result}) and the dimensionless constant $\alpha_\Lambda$ depends on the cut--off length (in relation to the Planck scale) and the effective number of fields involved. 

According to the Lifshitz theory developed in this paper, the vacuum energy and the associated cosmological constant are dynamical quantities\footnote{The theory of quintessence considers $\Lambda$ as a dynamical field as well, see e.g. Ref.~\cite{Caldwell}, but it does not include the physics of the quantum vacuum \cite{Milonni}. In this theory $\Lambda$ varies as well.}. The vacuum energy density responds to the evolving universe as described in Eqs.~(\ref{delta}) and (\ref{result}). From Friedman's Eq.~(\ref{f1full}) with Eq.~(\ref{state}) as equation of state follows
%%%%%%
\begin{equation}
\Lambda = \frac{8\pi G}{c^4} \,\epsilon_\Lambda = \frac{3}{c^2}\left(H^2 + \frac{\mathrm{k}\,c^2}{n^2a^2} +\frac{2}{3}\alpha_\Lambda\Delta - \epsilon_\mathrm{m}\right) \,.
\end{equation}
%%%%%'
In turn, the energy density of the quantum vacuum acts on the evolution of the universe as described in the equation of motion, Eq.~(\ref{cosmicmotion}). The vacuum energy appears as a correcting force to deviations from exponential expansion according to Eqs.~(\ref{correction}) and (\ref{oscillator}). Pure exponential expansion in flat space does not require any quantum correction. 

The theory does not predict a specific cosmological constant, as $\Lambda$ depends on dynamics, which implies that $\Lambda$ may have had different values. There are in fact two phases of cosmic expansion: one is measured \cite{Supernovae1,Supernovae2,CMBPlanck,Super} --- the recent phase, and one is conjectured --- the inflation \cite{Inflation} of the early universe where $H$ and hence $\Lambda$ was much larger. The theory of this paper accounts for both phases without requiring additional inflaton fields, although it is not yet clear how the inflationary phase ended and settled to the more sedentary pace of the recent era. 

This paper unifies for the first time the proven AMO physics of van der Waals and Casimir forces with the cosmological constant, following Zel'dovich's vision \cite{Zeldovich} with insights and tools from transformation optics \cite{LeoPhil} and modern quantum optics \cite{Grin1,Simpson,PXL,LeoSimpson,Grin2,Avni,KSW,PhilbinQ1,PhilbinQ2,Buhmann,Horsley,HorsleyPhilbin}. The theory still depends on one parameter  --- one may say one constant is traded for another --- but it also includes variations of the cosmological constant \cite{Super} and inflation \cite{Inflation}. Moreover, the parameter $\alpha_\Lambda$ has a physical meaning: it is given by the characteristic length near the Planck scale where the equivalence principle ceases to hold, and the effective number of fields involved. Its precise value cannot be predicted at present, but astronomical observations may infer $\alpha_\Lambda$ by fitting the measured expansion of the universe to the equation of motion, Eq.~(\ref{cosmicmotion}). Observations of the universe on the largest scale may thus probe the smallest scale of Nature. 

\section*{Acknowledgements}

This work was inspired by curious and courageous students (in alphabetical order):
Yael Avni,
Nimrod Nir,
Itay Griniasty,
Sahar Sahebdivan,
and
William Simpson.
I also thank Mikhail Isachenkov, Ephraim Shahmoon, and Anna and Yana Zilberg for discussions, support and inspiration. 
The European Research Council, the Israel Science Foundation, and the Murray B. Koffler Professorial Chair supported the work financially. 

\section*{Appendix A: Light in de Sitter space}

\renewcommand{\theequation}{A\arabic{equation}}
\setcounter{equation}{0}

The figures in Sec.~1 illustrate the propagation of light in the expanding universe, the role of cosmological horizons and the methodology of this paper. This appendix assembles the mathematical expressions applied there. 

Consider the simplest realistic model of the expanding universe: de Sitter space \cite{Peacock,Harrison,deSitter}. In the de Sitter universe, space is flat and expands with Hubble constant $H$ that is in fact constant. The solution of Eq.~(\ref{hubble}) for the Hubble constant is the exponential expansion:
%%%%%%
\begin{equation}
n=\mathrm{e}^{H t}
\label{ndS}
\end{equation}
%%%%%'
where the zero of the cosmological time $t$ is taken at $n=1$. Equation (\ref{tau}) gives the conformal time:
%%%%%%
\begin{equation}
\tau=-\frac{1}{H}\,\mathrm{e}^{-H t} = -\frac{1}{Hn}
\label{taudS}
\end{equation}
%%%%%'
with integration constant chosen such that $\tau$ runs along the negative axis from $\tau=-\infty$ for $t=-\infty$ to $\tau=0$ for $t=+\infty$ (Fig.~\ref{horizon}). Inverting Eq.~(\ref{taudS}) one obtains:
%%%%%%
\begin{equation}
t=-\frac{1}{H}\ln(-H\tau) \,.
\label{tdS}
\end{equation}
%%%%%'
Expressed in conformal time, space--time is conformally flat, Eq.~(\ref{conformallyflat}), such that light propagates like in free space, but this space--time is cut in half at $\tau=0$. Only the lower half corresponds to physical space--time. 

Cosmological horizons appear when the expansion velocity $v$ reaches $c$ where $v$ is defined in the fixed coordinates of Eq.~(\ref{fix}) and given by 
Eq.~(\ref{hubbleflow}). One obtains the horizon radius $\mathrm{r}_\mathrm{H}$ in fixed and $r_\mathrm{H}$ in co--moving coordinates:
%%%%%%
\begin{equation}
\mathrm{r}_\mathrm{H} = \frac{c}{H} \,,\quad r_\mathrm{H} = \frac{c}{nH} \,.
\end{equation}
%%%%%'
In fixed coordinates, the de Sitter horizon is stationary while space is expanding (Fig.~\ref{expansion}a) while in co--moving coordinates the horizon narrows exponentially in time $t$ (Fig.~\ref{expansion}b).

Consider the emission and reception of light at $\bf{r}=0$ surrounded by the cosmological horizon. As space--time is conformally flat in conformal coordinates, employ simply the flat--space Green functions $G_\pm$ of Eq.~(\ref{green0}) and express them in fixed coordinates:
%%%%%%
\begin{equation}
G_\pm = - \frac{n}{4\pi \mathrm{r}} \, \delta (\sigma \mp \rho)  \,,\quad  \sigma=\tau-\tau_0 \,,\quad \rho=\frac{\mathrm{r}}{n c} \,.
\label{greenfixed}
\end{equation}
%%%%%%
Here $G_+$ describes the emitted and $G_-$ the received light. Let me decompose $G_\pm$ into frequencies $\omega$ with respect to cosmological time $t$ by Fourier transformation: 
%%%%%%
\begin{eqnarray}
\widetilde{G}_\pm &\equiv& \int_{-\infty}^{+\infty} G_\pm\, \mathrm{e}^{\mathrm{i}\omega t}\,\mathrm{d}t \\
&=& -\frac{1}{4\pi\mathrm{r}} \int_{-\infty}^0 \delta\left[\tau_0(\tau/\tau_\pm-1)\right] \mathrm{e}^{\mathrm{i}\omega t}\,n^2\, \mathrm{d}\tau
\label{fourierdS}
\end{eqnarray}
%%%%%'
having changed the integration variable to $\tau$ according to definition (\ref{tau}) and used the abbreviation
%%%%%%
\begin{equation}
\tau_\pm= \tau_0\left(1\pm\frac{H\mathrm{r}}{c}\right)^{-1} .
\label{taupmdS}
\end{equation}
%%%%%'
At $\tau_\pm$ the delta function in Eq.~(\ref{fourierdS}) contributes to the integral. For the received light, $\widetilde{G}_-$ vanishes outside the cosmological horizon (for $\mathrm{r}>\mathrm{r}_\mathrm{H}=c/H$) as $\tau_->0$ there (since $\tau_0<0$). Nothing from beyond the horizon can be received. Put for simplicity 
%%%%%%
\begin{equation}
\tau_0=-\frac{1}{H} \,.
\label{tau0dS}
\end{equation}
%%%%%'
Write $\widetilde{G}_\pm$ in terms of the Heaviside step function $\Theta(\tau)$ as
%%%%%%
\begin{equation}
\widetilde{G}_- = \Theta(c-H\mathrm{r})\,{\cal G}_- \,,\quad \widetilde{G}_+ = {\cal G}_+ \,.
\label{calGdS}
\end{equation}
%%%%%'
In Eq.~(\ref{fourierdS}) express $\mathrm{e}^{\mathrm{i}\omega t}$ as $n^{\mathrm{i}\omega/H}$ according to Eq.~(\ref{ndS}) and $n$ as $(-H\tau)^{-1}$ according to Eq.~(\ref{taudS}), and finally $(-H\tau)^{-\mathrm{i}\omega/H}$ as $\exp[-\mathrm{i}(\omega/H)\ln (-H\tau)]$. This, with definitions (\ref{taupmdS}-\ref{calGdS}), gives the result:
%%%%%%
\begin{equation}
{\cal G}_\pm = -\frac{1}{4\pi\mathrm{r}} \left(1\pm\frac{H\mathrm{r}}{c}\right) \exp\left[\mathrm{i}\frac{\omega}{H} \ln  \left(1\pm\frac{H\mathrm{r}}{c}\right) \right] .
\label{resultdS}
\end{equation}
%%%%%'
For illustrating light propagation from the cosmological horizon in the expanding universe, Fig.~\ref{expansion}a shows $\mathrm{Im}{\cal G}_-$ (that is not singular at $\mathrm{r}=0$). In the vicinity of the horizon the phase of the wave diverges logarithmically, as it struggles to get away from the expanding space. Figure~\ref{horizon} shows the phase fronts $\varphi_\pm$ of the monochromatic waves $\widetilde{G}_\pm \mathrm{e}^{-\mathrm{i}\omega t}$ plotted as functions of conformal time $\tau$ and co--moving coordinates $\bm{r}$. One obtains from Eqs.~(\ref{tdS}) and (\ref{resultdS}) with $H\mathrm{r}=-r/\tau$ the phase
%%%%%%
\begin{equation}
\varphi_\pm = \frac{\omega}{H}\ln\left[H\left(\pm\frac{r}{c}-\tau\right)\right] \,.
\end{equation}
%%%%%'
This logarithmic phase is completely analogous to the one at the event horizon of the black hole \cite{Hawking1,Hawking2,Brout} and the Rindler horizon behind the Unruh effect \cite{Brout,Classical,Fulling,Davies,Unruh,Takagi}. It is an essential ingredient of particle creation at horizons, in particular of the Gibbons--Hawking effect \cite{GibbonsHawking} applied in this paper. Figure~\ref{method} shows the intensity profile of the dissipation $2c\widetilde{\Gamma}=\widetilde{G}_+ - \widetilde{G}_-$ in conformal coordinates $\tau$ and $\bm{r}$, using Eqs.~(\ref{calGdS}) and (\ref{resultdS}) with $H\mathrm{r}=-r/\tau$. This illustrates the quantity appearing in the fluctuation--dissipation theorem.

\section*{Appendix B: Green tensor}

\renewcommand{\theequation}{B\arabic{equation}}
\setcounter{equation}{0}

In the case of spatial curvature, the reduction of the electromagnetic field to two polarizations is not as straightforward as in flat space. This Appendix starts from the Green bi--tensor \cite{LeoPhil} of the field in homogeneous and isotropic space, and shows that the electric energy is reduced to an expression involving the scalar Green function with conformally--coupled wave equation, Eq.~(\ref{greencurveq}). The scalar Green function describes both polarizations, because they are equal in homogeneous and isotropic space. As the electric energy is equal to the magnetic energy \cite{LeoSimpson}, the derived expression accounts for exactly half the energy density. 

Consider the Green bi--tensor $\bf{G}$ {\cite{LeoPhil} in the conformal time defined in Eq.~(\ref{tau}). As Maxwell's equations are conformally invariant, the wave equation of $\bf{G}$ does not depend on the expansion factor $n(t)$, but only on the spatial profile $\nu(r)$. One obtains from the canonical commutation relation (\ref{commutator}) the propagation equation of $\bf{G}$ and Fourier--transforms with respect to conformal time: 
%%%%%%
\begin{equation}
\nabla\times\frac{1}{\nu}\nabla\times\widetilde{\bf{G}} - \nu k^2\, \widetilde{\bf{G}} = \delta(\bm{r}_1-\bm{r}_0) \mathbb{1}
\label{tensorwave}
\end{equation}
%%%%%'
with wavenumber $k=\omega/c$ and frequency $\omega$. Note that the transversality of the delta function \cite{LeoBook,MandelWolf} in the commutation relation (\ref{commutator}) cancels in $2c\bf{\Gamma}=\bf{G}_+-\bf{G}_-$ that enters the fluctuation--dissipation theorem of Secs.~2.4 and 2.5. Hence one can define $\bf{G}$ with $\delta(\bm{r}_1-\bm{r}_0) \mathbb{1}$ on the right--hand side instead of $\delta^\mathrm{T}(\bm{r}_1-\bm{r}_0)$, as done here.

The solution of Eq.~(\ref{tensorwave}) was found \cite{LPfish} in connection with perfect imaging \cite{LeoPhil} and is given by the expression \cite{LPfish}:
%%%%%%
\begin{equation}
\widetilde{G}_{ab}= \sum_{c\,d\,e f} \frac{[a\,c\,d]\,[b\,e f]}{\nu(r_1)\nu(r_0)k^2}\,\frac{\partial^2\nu(r')}{\partial x_1^c\,\partial x_0^e} \,\frac{\partial^2\widetilde{G}_\pm(r')}{\partial x_1^d\,\partial x_0^f}
\label{gleophil}
\end{equation}
%%%%%'
apart from a contact term proportional to $\delta(\bm{r}_1-\bm{r}_0) \mathbb{1}$ that does not contribute to the energy in the point--splitting method. All indices run in $\{1,2,3\}$, $\nu(r)$ is given by Eq.~(\ref{fish}), $r'$ by Eq.~(\ref{rprime}), $G_\pm$ denote the retarded and advanced scalar Green functions of Eq.~(\ref{greencurv}) and $[a\,b\,c]$ is the completely antisymmetric symbol in three dimensions \cite{LeoPhil} (encoding curls). 

The Green bi--tensor $G_{ab}$ describes the vector potential $\bm{A}$ in Cartesian $a$--coordinate emitted or received by a dipole of unity strength pointing in $b$--direction. As the electric field $\bm{E}$ is given by $-\partial_t\bm{A}$ in Coulomb gauge, the Fourier--transform of the correlation function $\langle \bm{E}(\bm{r}_1,\tau_1)\cdot \bm{E}(\bm{r}_0,\tau_0)\rangle$ with respect to $\sigma=\tau_1-\tau_0$ is proportional to $k^2\sum_a \widetilde{G}_{aa}$. One applies the double vector product \cite{LeoPhil}
%%%%%%
\begin{equation}
\sum_a[a\,c\,d]\,[a\,e f] = \delta_{ce}\,\delta_{df}-\delta_{de}\,\delta_{cf} \,,
\end{equation}
%%%%%'
calculates for $\bm{r}_1\rightarrow\bm{r}_0$ the limit $\frac{\partial^2\nu(r')}{\partial x_1^c\,\partial x_0^e} = \nu^2\,\delta_{ce}$, and obtains from Eq.~(\ref{gleophil}):
%%%%%%
\begin{equation}
k^2\sum_a \widetilde{G}_{aa} \sim 2 (\nabla_1\cdot\nabla_0)\,\widetilde{G}_\pm(r') \quad\mbox{for}\quad\bm{r}_1\sim\bm{r_0} \,.
\end{equation}
%%%%%'
This relationship carries over to the time--dependent $\bf{G}$ and, via the fluctuation--dissipation theorem, to $\bf{K}$. From the corresponding relationship, and Eqs.~(\ref{tau}) and (\ref{fields}), follows for the electric contribution $u_\mathrm{ED}$ to the energy density of Eq.~(\ref{correlation}):
%%%%%%
\begin{equation}
u_\mathrm{ED} = \frac{\hbar c \nu}{n}\,\nabla_1\cdot\nabla_0 K \,.
\end{equation}
%%%%%'
As the magnetic and electric energy densities are equal \cite{LeoSimpson} the total $u_\mathrm{vac}$ is twice $u_\mathrm{ED}$, which establishes Eq.~(\ref{uvacc}).

%%%%%%%%%%%%%%%%%%%%%%%%%%%%%%%%%%%%

%%%%%%%%%%%%%%%%%%%%%%%%%%%%%%%%%%%%%%%%%%%%%%%%%%%%%%%%%%%
\end{document}